\documentclass{aa}
\usepackage{graphicx}
\usepackage{txfonts}
\usepackage{natbib}
\usepackage{afterpage} % GDK added to make figure only pages
\graphicspath{{figures/}}
\begin{document}

   \title{Signatures of multiple episodes of AGN activity \\ in the core of Abell 1795}
   %\subtitle{}   
   %\titlerunning{}

   \author{G. Kokotanekov
          \inst{1}
          \and
          M. W. Wise \inst{2,1} 
          \and
          M.~ de Vries \inst{1}
          \and
          H.~T. Intema \inst{3}
          }

   \institute{Anton Pannekoek Institute for Astronomy, University of Amsterdam, Postbus 94249, 1090 GE Amsterdam, The Netherlands\\
              \email{g.d.kokotanekov@uva.nl}
         \and
             Netherlands Institute for Radio Astronomy (ASTRON), Postbus 2, 7990 AA Dwingeloo, The Netherlands\\
             \email{wise@astron.nl}
           \and
          Leiden Observatory, Leiden University, Niels Bohrweg 2, NL-2333CA, Leiden, The Netherlands
             }

   \date{Received March 29, 2018; accepted March 30, 2018}

%\abstract{}{}{}{}{} 
% 5 {} token are mandatory

 \abstract{
In this paper we analyze AGN activity signatures in the rich nearby galaxy cluster Abell 1795 aiming to confirm and characterize the long-term feedback history in the system.
We combine radio observations at 610 and 235~MHz from the Giant Metrewave Radio Telescope (GMRT) with 3.4 Msec X-ray data from the \textit{Chandra} Observatory.
Extracting radial temperature profiles, as well as X-ray and radio surface brightness profiles in three directions showing major morphological disturbances, we highlight the signatures of activity in the system.
For the first time we observe radio emission corresponding to the NW X-ray depression, which provides evidence in favor of the classification of the depression as a cavity.
We identify two other X-ray cavities situated NW and SW of the AGN.
While the central radio emission corresponding to the inner cavities shows a flatter spectral index, the radio extensions associated with the farthest X-ray cavities consist of aged plasma. 
All observed signatures both in radio and X-ray are consistent with several consecutive episodes of AGN activity, which gave rise to the observed morphology NW and SW from the core. 
In the southern region, we confirm the cooling wake hypothesis for the origin of the long tail.
The deep X-ray data also allows us to distinguish significant distortions in the tail's inner parts, which we attribute to the activity of the AGN.
}

 \keywords{galaxy clusters, X-ray cavities, AGN feedback, radio continuum, galaxies: clusters: 
           general —    galaxies: clusters: individual A1795 — galaxies: clusters: 
           intracluster medium — radio continuum: general — X–rays:galaxies: clusters}

 \maketitle
%
%-------------------------------------------------------------------

%%%%%%%%%%%%%%%%%%%%%%%%%%%%%%%%%%%%%%%%%%%%%
%% INTRODUCTION
%%%%%%%%%%%%%%%%%%%%%%%%%%%%%%%%%%%%%%%%%%%%%

\section{Introduction}

Central active galactic nuclei (AGN) play a crucial role in the evolution of galaxy cluster cores providing energy that regulates both the growth of the black hole and the formation of stars in the surrounding galaxy \citep{Galaxy_Formation_Silk, MBH_Msigma_Gebhardt, SMBH_Galaxy_Ferrarese}. In the classical model, symmetric cavities observed in X-rays are blown into the ICM on either side of the brightest cluster galaxy (BCG) by the expanding radio lobes generated by the central AGN. For many systems, these X-ray cavities are observed to be filled with radio emitting plasma \citep{Fabian2000, McNamara2000}, further supporting this model. Deeper X-ray observations and new radio data, however, show that establishing this simple correspondence over longer feedback timescales can be more difficult.

Deep X-ray observations of the cores of nearby relaxed galaxy clusters have revealed a wide range of irregular surface brightness features not easily identified as simple, spherical cavities  (e.g., Perseus: \citealp{Boehringer1993, Fabian2006, Fabian2011, Zhuravleva2015} and Hydra A: \citealp{McNamara2000, Nulsen2002, Wise2007}). In a number of these systems,  low-frequency radio observations have shown diffuse, extended steep-spectrum emission presumably representing the signatures from older AGN outbursts (M87: \citealp{DeGasperin2012}, Perseus: \citealp{Burns1992, Sijbring1993}, Hydra A: \citealp{Wise2007}, 2A 0335+096: \citealp{GDK2017}). At radio frequencies below 1 GHz, however, the close morphological correspondence between the observed X-ray structures and radio emission seen at higher frequencies is not always apparent. To understand the impact of feedback over the lifetime of the cluster, it is important to determine how these complicated morphologies map to the long-term AGN activity in the core.

Abell 1795 (hereafter A1795) is an excellent example of a source exhibiting such complicated structures. It is a rich, nearby cluster of galaxies ($z = 0.063$, with an angular scale of 1.22 kpc per arcsec) that hosts the powerful Fanaroff--Riley type I (FR I) radio source 4C\,26.42. While classified as a relaxed, cool-core cluster because of its regular morphology on larger scales \citep{Buote1996}, A1795 shows evidence of a variety of activity throughout the cluster core. Deep \textit{Chandra} data have revealed a myriad of structures \citep{Markevitch2001, Ettori2002, Walker2014, Crawford2005, Ehlert2015, Walker2017} not easily described by a simple feedback scenario involving only a single outburst.

One such structure, a large $r\sim$34\,kpc depression to the northwest of the central AGN, was previously identified in \textit{Chandra} data \citep{Crawford2005} and classified as a cavity by \cite{Walker2014} based on its morphology. Despite its suggestive X-ray properties, however, previous studies at higher frequencies have failed to detect radio emission associated with the cavity. The cavity also lacks a counterpart on the opposite side of the cluster as well as other typical observational signatures normally associated with X-ray cavities, such as evidence of abundance gradients or optical filaments created as the cavity rises in the cluster atmosphere. Taken together, this lack of additional evidence has made it difficult to definitively classify the NW structure as a traditional cavity associated with an outburst.

The core of A1795 also contains a dramatic filament of cool gas extending over $\sim$50\,kpc to the south of the central cD galaxy \citep{Fabian2001} and coincident with an optical H$\alpha$ filament \citep{Cowie1983}. Several theories for the origin of this filament have been proposed, but \cite{Fabian2001} have argued that the most likely origin is that the feature represents a ``cooling wake'' produced as the ICM cools behind the central galaxy as it moves relative to the surrounding gas. Alternatively, this filament may consist of cold gas from the central galaxy which has been stripped out by dynamical friction due to the motion of the cD galaxy. Regardless of the exact formation mechanism, the presence of such a dramatic feature and its implications for the relative motions of the gas and central galaxy clearly point to a more complicated evolution of the core region in A1795 than can be explained by a simple AGN outburst model.

Finally, \cite{Markevitch2001} have shown that the inner 60 kpc core of A1795 exhibits a surface brightness edge, which they interpret as a ``cold front'' or contact discontinuity caused by cool gas in the core sloshing in the central potential well of the cluster. In this scenario, the cool gas interior to the surface brightness edge is believed to be sloshing roughly north to south and is observed at or near the point of maximum displacement. The observed filament of cool material is then a cooling wake produced, not by motions of the cD galaxy in the potential well itself, but rather by the movement of the sloshing gas. The cumulative impact of the larger scale sloshing motion of the cool core and AGN activity \citep{Ehlert2011, Blanton2011} is evoked to explain the profoundly disturbed morphology observed in the cool core of A1795.

Altogether these features imply a dynamic environment in the core of A1795 over the last several hundred Myrs, and one that does not easily conform to the simple picture of an AGN-driven outburst. In this paper, we present a joint analysis for the core of A1795 combining a deep, 3.4\,Msec archival X-ray dataset from \textit{Chandra} and new low-frequency GMRT observations at 235 and 610\,MHz. We describe the X-ray and radio data along with the data reduction process in Sect.\,\ref{secP2:data}. Sections\,\ref{secP2:xray_morphology} and \ref{secP2:radio_morphology} compare the morphologies of the different features revealed in the deep \textit{Chandra} X-ray maps with the low-frequency radio structures observed in the core of A1795. We perform a multiwavelength sector analysis in Sect.\,\ref{secP2:cavities} to isolate various features of interest in the core and examine the underlying thermodynamic properties of the gas in an effort to constrain the integrated feedback history over the last few hundred Myr. The properties of the cold filament are presented in Sect.\,\ref{secP2:tail} in the context of current theories for its origin. The paper concludes in Sect.\,\ref{secP2:conclussions} with a summary of our analysis and a discussion of the implications of our results.

%%%%%%%%%%%%%%%%%%%%%%%%%%%%%%%%%%%%%%%%%%%%%
%% DATA PROCESSING
%%%%%%%%%%%%%%%%%%%%%%%%%%%%%%%%%%%%%%%%%%%%%

\section{Data reduction}
\label{secP2:data}      
        
\subsection{X-ray data}
\label{secP2:data_xray} 

We have extracted and reprocessed existing archival \textit{Chandra} data for A1795, which has been used extensively as a \textit{Chandra} calibration and reference target since launch and, consequently, has one of the deepest accumulated exposures of any cluster target. In this work, we analyze over 3.4\,Msec of \textit{Chandra}  data, consisting of 148 individual ObsIDs taken between December 1999 and November 2016. Most of these data were observed as part of the \textit{Chandra} calibration program and include 33 and 115 pointings taken with the ACIS-S and ACIS-I detectors, respectively. In order to assess the calibration properties of individual ACIS CCDs, the aimpoints for many of these observations also vary. When combined, these variations result in a somewhat broader point spread function (PSF) across the core of A1795. Our features of interest are significantly larger than the \textit{Chandra} PSF, so this broadening should have no appreciable effect on the analysis presented here. 

Each ObsID was reprocessed using \texttt{CIAO}\,4.8 \citep{CIAO2006} and \texttt{CALDB}\,4.7.2 to apply the latest gain and other calibration corrections. Standard filtering was applied to the event files to remove bad grades and pixels. Additionally, all ObsIDs were examined for contamination due to strong background flares and filtered to remove the effected time intervals using the \texttt{CIAO} tool \texttt{deflare}. These filtered event files were also  corrected  for the presence of ``out-of-time'' events detected during the frame readout process using the \texttt{readout\_bkg} routine. The final combined exposure time after all corrections was 3.45\,Msec. 

ACIS blank-sky event files from the \texttt{CALDB} were used to construct background event files for each ObsID. The blank-sky event files were reprocessed in a manner similar to the observation data to apply the latest calibrations. The \texttt{CALDB} ACIS blank-sky event files have already been screened for background flares so no additional filtering for flares was applied. Finally, the exposure times for each blank-sky event file were scaled to match the observed background rate in the $10-12$ keV band for each ObsID. These scaled background event files were used in all subsequent spectral analyses discussed below.

\subsubsection{Surface brightness maps} 

An X-ray mosaic of the surface brightness in A1795 was constructed by reprojecting all event files to a common tangent point on the sky before combining them. The event file for each ObsID was reprojected to match the coordinate frame of the reference ObsID 493. To correct for variations in the exposure and other instrumental artifacts across the field, instrument and exposure maps were constructed for each ObsID in the $0.5-7.0$ keV energy range. We have accounted for the energy dependence of the instrument response by utilizing a spectral weighting determined by fitting the integrated spectrum for the central $2^{\prime}$ region of A1795. Bright point sources were excised from the region and the resulting total spectrum was fit with a single temperature, \texttt{APEC} \citep{Smith2001} thermal model multiplied by foreground Galactic absorption. The Galactic absorption was fixed to a value of $N_H = 1.2 \times 10^{20}$ as determined by the LAB Survey of Galactic HI \citep{Kalberla05}. This model gives a good fit to the data with a reduced $\chi^2 = 1.1$ and best-fit temperature and metallicity of $kT = 5.7$ keV and $Z = 0.41$, respectively. 

\begin{figure}[t]
\begin{center}
\includegraphics[height=2.9in]{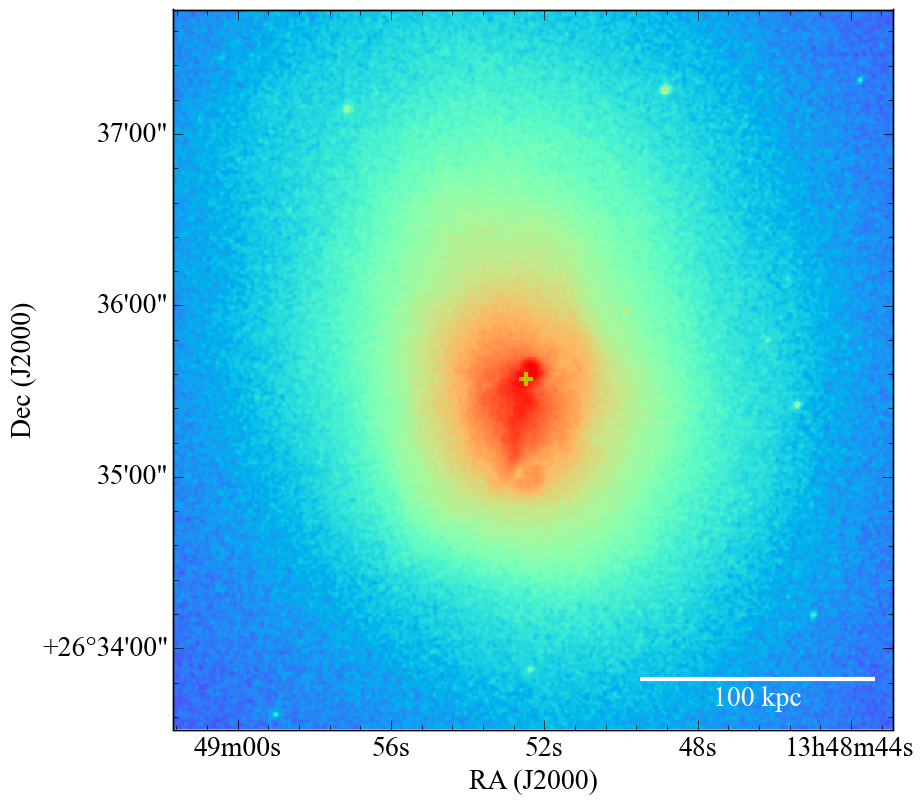}
\end{center}
\caption{\small Exposure-corrected, background-subtracted 0.5--7.0 keV surface brightness mosaic for the core of A1795 constructed from the existing 3.4\,Msec \textit{Chandra} data. The field measures $4 \times 4$ arcmin$^2$ and has been smoothed with a $\sigma = 3\arcsec$ Gaussian. The cross indicates the position of the center of the radio source.
\label{figP2:X-ray_surf_brightness}}
\vspace{0.15in}
\end{figure}

The resulting spectrally weighted exposure maps were combined into a single mosaic after reprojection to the common tangent point. Similarly, a map of the background surface brightness was constructed by reprojecting the scaled, blank-sky event files for each ObsID and combining them. To obtain the final, flat-fielded, surface brightness map, we subtracted the total background map from the source counts mosaic and divided by the exposure map for the field. The resulting background-subtracted, exposure-corrected surface brightness mosaic for the core of A1795 in the energy range $0.5-7.0$ keV is shown in Fig.\,\ref{figP2:X-ray_surf_brightness}. Various features of interest are described in Sect.\,\ref{secP2:xray_morphology}.

%\begin{figure}[!htbp]

\begin{figure}[t]
\begin{center}
%% original size
\includegraphics[height=2.9in]{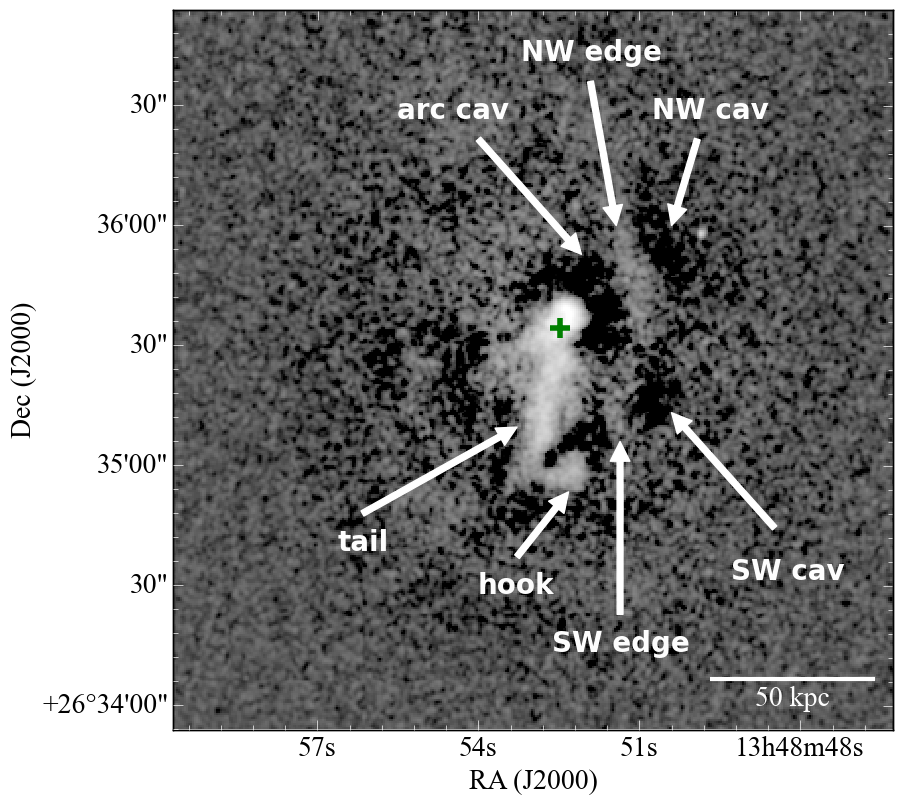}
\end{center}
\caption{\small \textit{Chandra} X-ray surface brightness residual map of A1795 in the 0.5--7 keV band. The image was produced by applying a radial unsharp masking (RUM) algorithm to the image in Fig.\,\ref{figP2:X-ray_surf_brightness} as described in the text. The cross indicates the position of the core of the radio source. Various features of interest in the core of A1795 are labeled. 
\label{figP2:X-ray_residual}}
\vspace{0.15in}
\end{figure}

Although several morphological features  like  the cool tail of gas to the south are visible in Fig.\,\ref{figP2:X-ray_surf_brightness}, the overall surface brightness distribution in the core of A1795 is quite smooth and other features are less apparent. To enhance the visibility of these other features, we  constructed an X-ray residual map using a radial unsharp masking (RUM) technique. This technique is described in \cite{Wise2018} and is similar to traditional unsharp masking, except rather than computing the difference between the original image when convolved with two different 2D smoothing kernels, the RUM algorithm applies a single 1D smoothing kernel radially from the image center. A Radon transform \citep{Radon1986} was applied to the original surface brightness map to convert from $(x,y)$ image coordinates to $(r, \Theta)$ coordinates, where $r=0$ is taken to be the center of the radio source. 
The resulting transformed image was then smoothed with a 1D Gaussian kernel with $\sigma = 8$\,arcsec along the radial direction. Finally, the smoothed image in $(r, \Theta)$ coordinates is transformed back to the $(x,y)$ image plane and subtracted from the original surface brightness map. This technique has the advantage of being less prone to creating circularly symmetric bowl features due to the mismatch between the two 2D smoothing kernels used in traditional unsharp masking techniques \citep{Wise2018}. The resulting residual image is shown in Fig. \ref{figP2:X-ray_residual}.

%\begin{figure}[!htbp]

\subsubsection{Spectral fitting}
\label{secP2:fitting}

In the subsequent analysis, we perform spatially resolved spectral fitting to determine the underlying thermodynamic properties of the gas in the core of A1795. We consider both 2D spectral mapping and radial profiles extracted in different sectors of interest as determined by the morphological features seen in the surface brightness maps. For the 2D spectral maps, we  used the \texttt{contbin} algorithm developed by \cite{Sanders2006} to define a set of extraction regions covering a $10\times10$\,arcmin$^2$ region centered position of the central AGN in A1795. A position for the AGN of (RA 13:48:52.473, Dec +26:35:34.385) was used based on the 1.4 GHz VLA map of \cite{Birzan2008}. The regions were defined to have a constant signal-to-noise ratio (S/N) of 100 after background subtraction, and therefore roughly the same errors on the derived fit parameters.

Radial profiles were constructed by defining a set of annular extraction regions centered on the position of the AGN and adaptively adjusted in width so as to contain a given S/N after background subtraction. Unless noted otherwise, a minimum S/N of 100 was used and annuli out to a radius of 5 arcmin were defined. We  defined annular extraction regions encompassing the entire cluster profile and a set of sector profiles limited to a given angular range and designed to isolate features of interest along a given radial direction through the core of A1795. These radial sectors and their correspondence to specific surface brightness features are discussed in Sect.\,\ref{secP2:cavities}.

The spectra in a given region were extracted using the \texttt{CIAO}\,4.8 tool \textit{specextract}. This tool produces source and background spectra, an instrument response (RMF) file, and an effective area (ARF) file, all of  which are used in the spectral fitting analysis. This set of spectral files is produced for every extraction region for each CCD from each ObsID. The resulting set of spectral files from all ObsIDs contributing to a given extraction region are then combined into a single composite set using the \textit{combine\_spectra} tool in the default way. After combination, we adjust the total exposure time of each extraction region to correct for the fact that regions that fall on different chips within the same ObsID are counted twice when \textit{combine\_spectra} adds them, which incorrectly increases the total exposure time of the combined spectrum. The final set of merged spectral files for all extraction regions are then used as input to the spectral fitting analysis.

The ObsIDs for A1795 used in our analysis span a wide range in time in which the effective area of ACIS has significantly changed, and also include observations taken with both the ACIS-S and the ACIS-I detectors. The time changes and the intrinsic differences of the two types of CCDs predominatly effect the spectral response below 2.0 keV. 
The exposure-weighted averaging scheme employed by the \texttt{CIAO} \textit{combine\_spectra} tool effectively averages over these variations to produce a single composite ARF and RMF for a given region. When performing spectral fits with these composite ARFs and RMFs, there is the risk of deriving incorrect or biased results if the variations from the average are too large. We note, however, that for the range of hotter gas temperatures ($3-6$ keV) in the core of A1795 these variations in the soft response of the telescope are expected to be less significant.

To check whether the composite ARFs and RMFs give results that are accurate enough, we separated the datasets into subsets of ObsIDs observed on ACIS-S and ACIS-I, respectively. These subsets were then each processed separately to produce summed spectra and composite ARFs and RMFs for ACIS-S and ACIS-I. The spectral analysis in the different sectors was then redone using a joint fit to the composite ACIS-S and ACIS-I spectra in each region. The resulting temperature profiles and other spectral fit parameters all agree with the original results based on the composite spectral files from all ObsIDs to within less than 5\%. Given that the typical measurement error on the derived spectral parameters is typically $5-10$\%, we conclude that the composite spectral files are sufficient to derive the underlying spectral properties of the ICM in A1795.

All spectral analyses presented were performed using \texttt{Sherpa} in \texttt{CIAO}\,4.8 over the $0.5-7.0$ keV energy range. Except where noted otherwise, all data were grouped to contain a minimum of 25 counts per bin and fit with a \texttt{PHABS*APEC} spectral model using the default $\chi^2$ fit statistic.

\subsection{Radio data}

\begin{figure*}[!htbp]
\begin{center}
\includegraphics[height=2.9in]{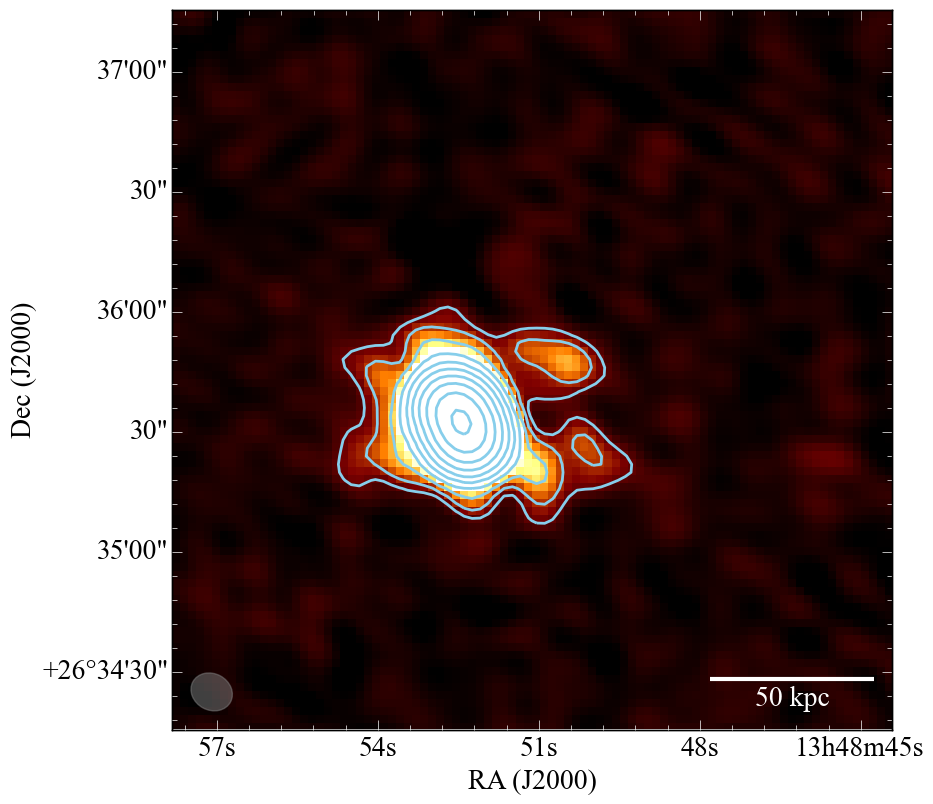}
\includegraphics[height=2.9in]{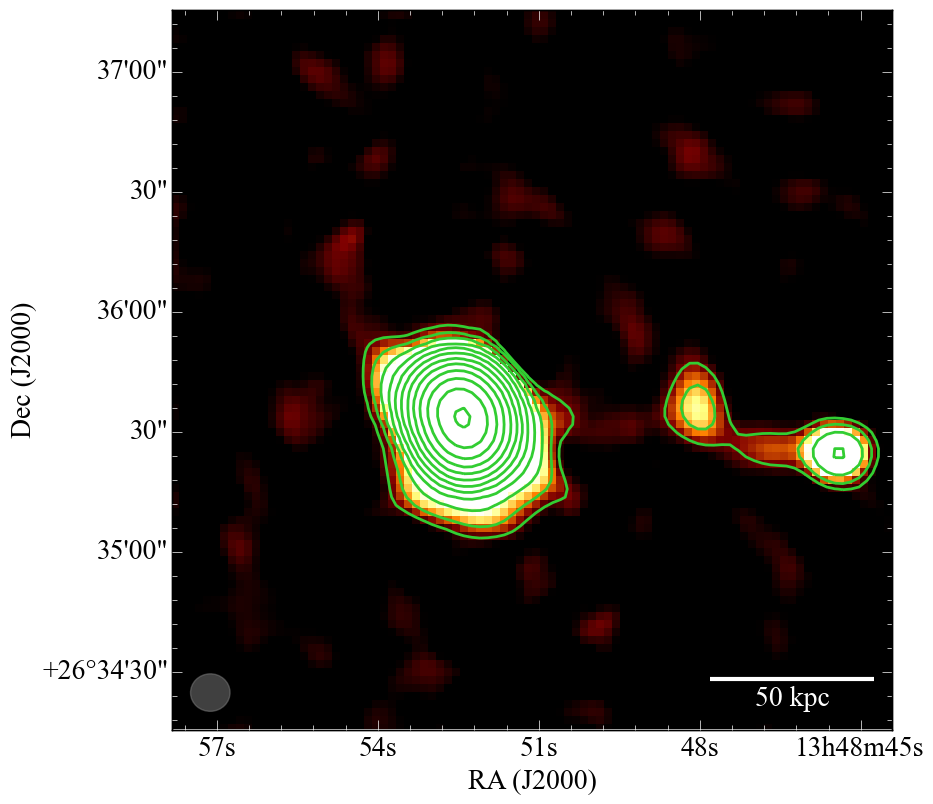}
\end{center}
\caption{\small 
GMRT maps of A1795.
\textit{Left}: Radio map at 235 MHz with resolution of $10.7 \times 9.1^{\prime\prime}$ and rms noise level of 0.8 mJy/beam. Contours are drawn at 5$\sigma$ $\times$ [1, 2, 4, 8, 16, 32, 64, 128, 256, 512]. 
\textit{Right:} Radio map at 610 MHz with resolution of $9.9 \times 9.4^{\prime\prime}$ and noise level of 0.18 mJy/beam. Contours are drawn at 3$\sigma$  $\times$ [1, 2, 4, 8, 16, 32, 64, 128, 256, 512, 1024, 2048].
\label{figP2:GMRT}}
\vspace{0.15in}
\end{figure*}

We processed archival radio observations from the Giant Metrewave Radio Telescope \citep[GMRT;][]{Swarup1991} on A1795 using the fully automated \texttt{SPAM} pipeline \citep{Intema2014}, which incorporates correction for direction-dependent effects \citep{Intema2009}. The observations under project code 20\_016 consist of simultaneous dual-frequency observations at 235 and 610 MHz, using bandwidths of 16.7 and 33.3 MHz, respectively. The total on-source time was 273~minutes (or 4.55~hours). Flux and bandpass calibrations were derived from 3C468.1 and 3C286, respectively, using flux models based on the work by \citet{Scaife2012}. Phase calibrations were not derived from the secondary calibrators, but instead from (self-)calibration against the TGSS \citep{Intema2017} in the case of 235 MHz, and against the 235 MHz image in the case of the 610 MHz. During the pipeline processing about 50\% of the data was removed because of non-functional antennas, radio-frequency interference (RFI), and bad ionospheric conditions. The output images and visibility data were post-processed to allow final imaging in \texttt{CASA}. For this, all sources outside a radius of $2^{\prime}$ from A1795 were subtracted from the visibility data, effectively removing all direction-dependent effects.
        
To image the GMRT data we used the interactive cleaning of \texttt{CASA} setting Briggs weighting scheme with a robust parameter of $-$0.2 for both frequencies. 
We imposed a Gaussian taper with outer $uv$-cut of 10$^{\prime\prime}$ at 610 MHz in order to highlight the large-scale  structures and achieve comparable resolution with the 235~MHz image. 
The end resolution of the 610 MHz image is $9.9 \times 9.4^{\prime\prime}$ at an rms noise level of 0.18 mJy/beam.
The 235 MHz image has a resolution of $10.7 \times 9.1^{\prime\prime}$ and a noise level of 0.8 mJy/beam.
The GMRT images are presented in Fig. \ref{figP2:GMRT}.

We complement our study with published VLA maps at 1.4~GHz in A-configuration \citep{Birzan2008} and C-configuration \citep{Giacintucci2014}.
The A-configuration image has an rms noise level of 70 $\mu$Jy/beam. It is based on an observation with exposure of 174 min and a bandwidth of 50 MHz.
The size of the FWHM of the synthesized beam is $1.2 \times 1.1^{\prime\prime}$.
The resolution of the C-configuration map is $19 \times 16^{\prime\prime}$.
The corresponding radio data was integrated for 130 min at a bandwidth of 6 MHz. The resulting noise level is 150 $\mu$Jy/beam.

%%%%%%%%%%%%%%%%%%%%%%%%%%%%%%%%%%%%%%%%%%%%%
%% X-RAY MORPHOLOGY
%%%%%%%%%%%%%%%%%%%%%%%%%%%%%%%%%%%%%%%%%%%%%

\section{X-ray morphology}
\label{secP2:xray_morphology}

In this section we discuss the observed X-ray morphology in the core of A1795. We identify the most pronounced features of interest, which we study in detail in the subsequent sections.

\subsection{Large-scale structure}

A1795 has been extensively observed  at X-ray wavelengths in the core and on larger scales.
It has been shown that the structure of the core of A1795 exhibits a very smooth and circularly symmetric surface brightness distribution at a radius beyond $\sim$80 kpc from the AGN \citep[e.g.,][]{Ettori2002, Ehlert2015}. In the surface brightness map presented in Fig.\,\ref{figP2:X-ray_surf_brightness}, we focus on the central region of A1795 within a radius of 150\,kpc centered on the central AGN. Inside a radius of $70-80$\,kpc, the structure of the ICM appears heavily disturbed. The cool tail is quite evident; it  extends due south away from the core and appears to end with a hook-like structure that bends toward the west. We are also able to identify the inner rim of the large depression NW from the core studied by \cite{Walker2014}. %

These features and others are even more prominent in the X-ray residual map presented in Fig.\,\ref{figP2:X-ray_residual}. In the NW direction we see a number of features including the large outer depression. Its inner rim is a distinct edge of bright emission situated 28~kpc NW from the core. The most pronounced region of the depression is situated at a distance of $\sim$ 40 kpc NW from the center and appears elongated in the tangential direction. The NW depression has been classified as a cavity and studied in detail by \cite{Walker2014}, who define its size based on the curvature of the concave NW edge. 
They measure a distance from the central galaxy to the center of the outer NW cavity of 54~kpc and derive an age of $t_s = 41$ Myr based on the sound speed.
Closer to the AGN we observe another depression surrounding the active core to the N, NW, and W. 
This depression appears severely distorted and is located $\sim$15~kpc from the center, being nested between the centrally peaked core of the X-ray surface brightness and the NW edge.

\begin{figure*}[!htbp]
\begin{center}
\includegraphics[height=2.8in]{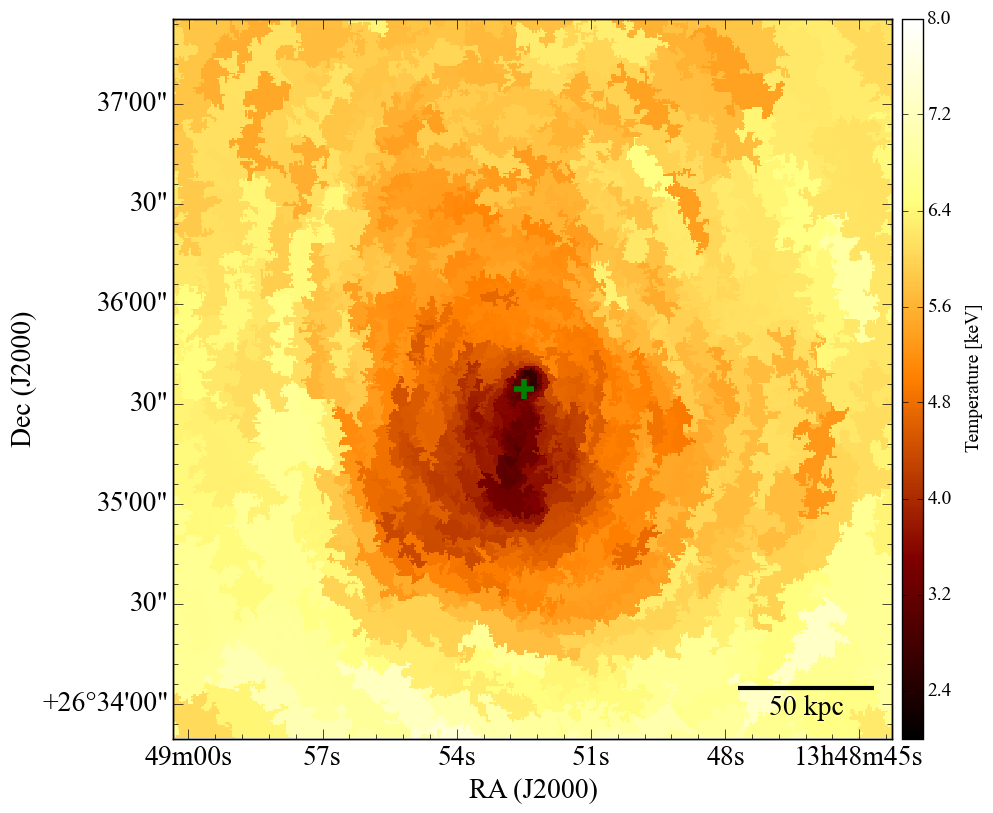}
\includegraphics[height=2.8in]{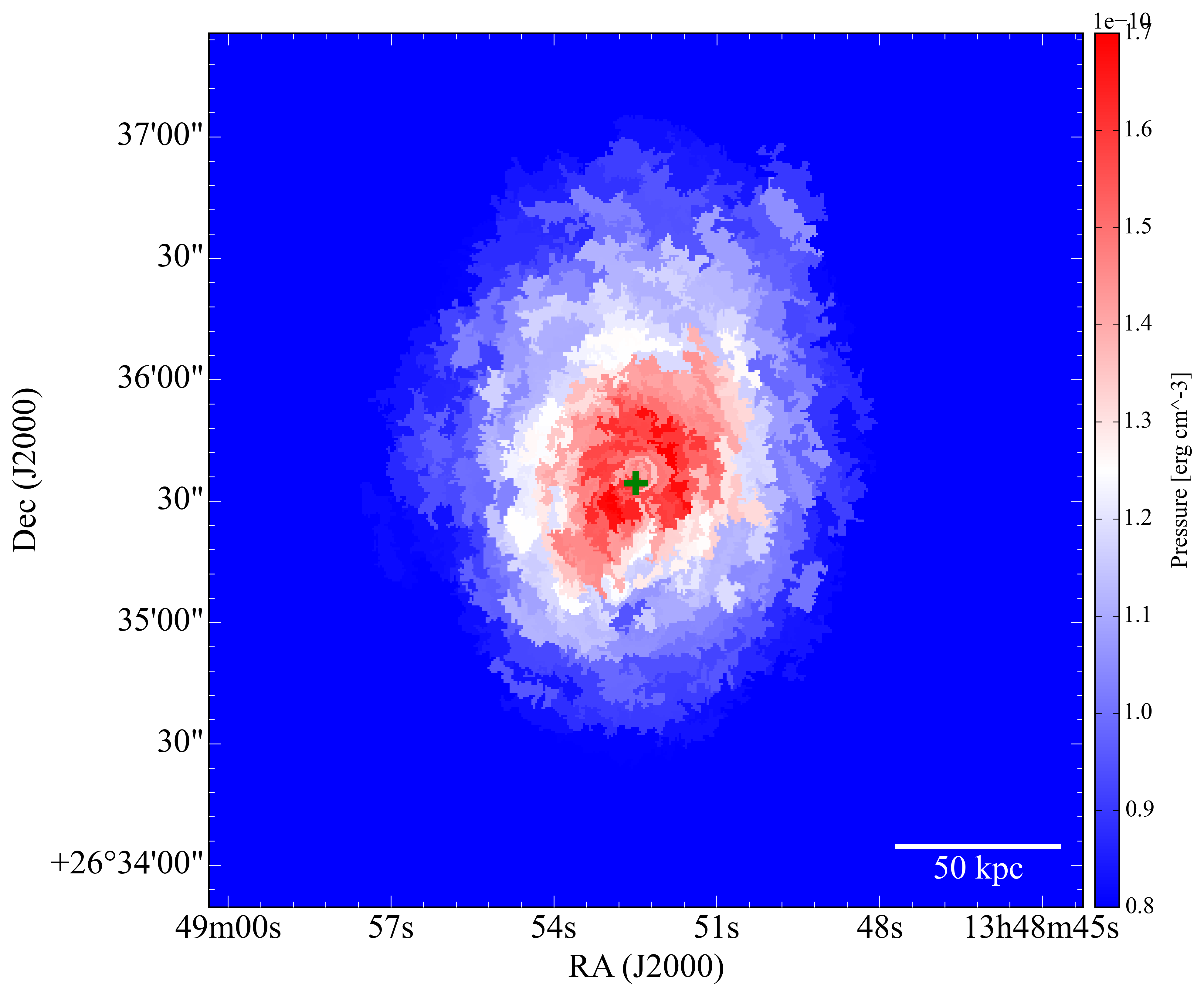}
\end{center}
\caption{\small Temperature (\textit{left}) and pressure (\textit{right}) maps of the core of A1795 within a radius of 130 kpc from the AGN.
Both maps are derived from \textit{Chandra} imaging X-ray spectroscopy using regions with a S/N of 100, as described in Sect. \ref{secP2:data_xray}. The cross indicates the center of the radio source.
\label{figP2:temp_pres_big}}
\vspace{0.15in}
\end{figure*}

We also identify similar features, both an edge and outer X-ray depression, to the SW.
The SW edge peaks 25 kpc from the BCG. Thus the SW edge appears to be situated slightly closer to the core than the NW edge. The depression is situated outside of the SW edge, at a distance 33~kpc from the AGN. 
These are new structures that have not been studied previously and, as we show in the following section, they are not artifacts from the construction of the residual map. We discuss these features in light of a possible feedback origin in Sect.\,\ref{secP2:cavities}.

The X-ray residual map clearly shows the cold tail to the south and confirms the finding from previous studies \citep[e.g.,][]{Ehlert2015} that the cold tail is very confined.
It stretches as far as 47 kpc from the core, ending with a hook extending to the west. The new \textit{Chandra} data, however, allow us to examine the substructure of the tail in greater detail than before. Our X-ray residual map reveals small-scale inhomogeneities in its structures and lets us identify kinks in its direction of propagation. The properties of the cool tail are discussed in more detail in Sect.\,\ref{secP2:tail}.

\cite{Crawford2005} identified a small-scale hole in the X-ray emission locked between the hook and the end of the tail, centered at RA 13:48:52.7, Dec +26:35:01.8.
Later this depression was analyzed by \cite{Walker2014} in an attempt to identify its origin. 
Our residual map reveals a continuation of this depression to the north, which is enclosed between the hook, the tail, and the SW edge. We defer the detailed analysis of the resulting irregularly shaped depression to future studies.

Although the eastern half of the X-ray map seems less rich in structure than the western side, the emission in this region is far from uniform. Our map reveals several ripples and irregularly shaped depressions in the region E and SE from the center. However, the observed features appear severely distorted and are not as easy to classify. Since these structures are not as distinct and we do not have enough evidence to identify their nature, we leave their analysis for future work.

\subsection{Temperature and pressure}

If the morphological features identified in the residual map are related to AGN activity, then the thermodynamic properties of the gas in the core should show evidence of the energy deposited into the ICM by the activity. To look for this evidence, we  constructed temperature and pressure maps for the ICM in the center of A1795. As discussed in Sect.\,\ref{secP2:fitting}, we  defined an adaptively sized grid of 2D extraction regions over the central $10\times10$\,arcmin$^2$ of A1795 and extracted summed spectra for all ObsIDs contributing to a given region. 
These spectra were then fit with a \texttt{PHABS*APEC} model to determine the projected gas temperature as a function of position in the core. The Galactic absorption was held fixed to a value of $N_H = 1.2 \times 10^{20}$ and the abundance was a free parameter in the model. The normalization of the fitted spectrum can be used to derive the particle number density, $n$, and when combined with the temperature map to construct a projected map of the gas pressure in the core. The projected temperature and pressure maps are presented in Fig.\,\ref{figP2:temp_pres_big}.

The temperature map shows the NW X-ray depression as a region of enhanced temperature compared to the ambient. The most pronounced temperature difference between the depression and the surrounding medium appears at the position of the NW edge. The dense NW edge clearly separates the $\sim$ 6 keV gas within the cavity from the $\sim$ 5 keV gas between the edge and the active core. The abrupt NW temperature jump has been  analyzed by several authors \citep[e.g.,][]{Markevitch2001, Ehlert2015, Walker2014}. It has been shown that it is not consistent with a shock moving out since in this case the gas should be hotter behind the compressed region. The observed edge is more consistent with a cold front \citep{Markevitch2001}, due to its reversed temperature structure.

The overall large-scale temperature structure is consistent with previous publications.
We see the cold front situated $\sim$80 kpc south of the center reported by \cite{Markevitch2001}.
The cold tail is very pronounced in the temperature map appearing as a strip of material much colder than its surroundings.

The pressure morphology in the central $\sim$20 kpc from the AGN is radially symmetric, and in this region we measure systematically higher values than at the central AGN.
Outside this radius the pressure structure appears irregularly distributed.
The map reveals large-scale extension of high pressure relative to the ambient stretching SE and NW. 
This symmetry in the NW--SE direction might be evidence of an older symmetric outflow due to the traditionally expected bipolar AGN ejection.

%%%%%%%%%%%%%%%%%%%%%%%%%%%%%%%%%%%%%%%%%%%%%
%% RADIO MORPHOLOGY
%%%%%%%%%%%%%%%%%%%%%%%%%%%%%%%%%%%%%%%%%%%%%

\section{Radio morphology}
\label{secP2:radio_morphology}

In this section we discuss the structure in the core of A1795 observed at different radio frequencies. 
We describe the morphological features revealed by our new GMRT images at 235 and 610 MHz.
We also include two previously published VLA observations at 1.4 GHz. 
The high-resolution map of \cite{Birzan2008} shows the inner radio lobes, while the lower-resolution image of \cite{Giacintucci2014} reveals the large-scale structure of the source.
We discuss the radio lobes in the context of the pressure structure of the core, while the morphology observed in the map of \cite{Giacintucci2014} is compared with the structures observed with GMRT.

\subsection{VLA 1.4 GHz A-configuration}
\label{secP2:high-freq_Laura}

\begin{figure}[!htbp]
\begin{center}
\includegraphics[height=3.40in]{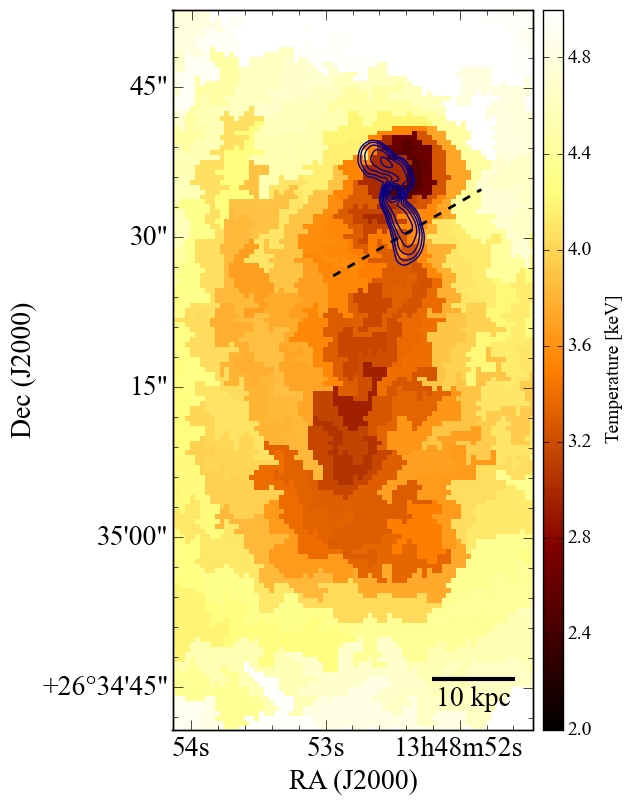}
\end{center}
\caption{\small Temperature map of the cold tail with radio contours at 1.4 GHz from the VLA in A-configuration from \cite{Birzan2008}. Contours are drawn at 10~mJy/beam $\times$ [1,1.4,2,2.8,4,5.7,8,11]. 
The black dashed line indicates the division between  the inner and outer tail, as defined in Sect. \ref{secP2:tail}.
\label{figP2:temp_tail}}
\vspace{0.15in}
\end{figure}

The inner radio lobes of the central AGN have been  studied using VLA observations at 8.2 GHz and 1.4 GHz \citep[e.g.,][]{GeOwen1993, Birzan2008}.
In Fig. \ref{figP2:temp_tail} we present the radio contours of the A-configuration 1.4 GHz image from \cite{Birzan2008} overplotted on an enlarged section of our temperature map showing the active core and the cool tail. 
The observed high-frequency radio emission is confined to within $\sim$10$^{\prime\prime}$  of the active nucleus.
The high-frequency observations reveal distorted structure with two lobes bent through 90\degr within $\sim$2$^{\prime\prime}$ of the nucleus. 

The northern lobe appears more affected by the movement of the core.
Figure \ref{figP2:temp_tail} shows that the N lobe is well confined by an arc-shaped structure with temperature of 2.5~keV. 
This arc-like structure has been interpreted by \cite{Ehlert2015} as consistent with a cold front due to movement of the BCG northward.
Similarly to the northern lobe, the southern lobe changes direction from SE to SW when it meets a region of higher pressure at around 1.5$^{\prime\prime}$ to the SE of the active nucleus.
After its deflection, the southern radio lobe flows through a channel of reduced pressure, which leads to the western part of the tail (see Sect. \ref{secP2:tail}).
The channel of lower pressure navigates the lobe toward the SW and allows it to expand in this direction.

\begin{figure*}[!htbp]
\begin{center}
%% 2 column
\includegraphics[height=2.90in]{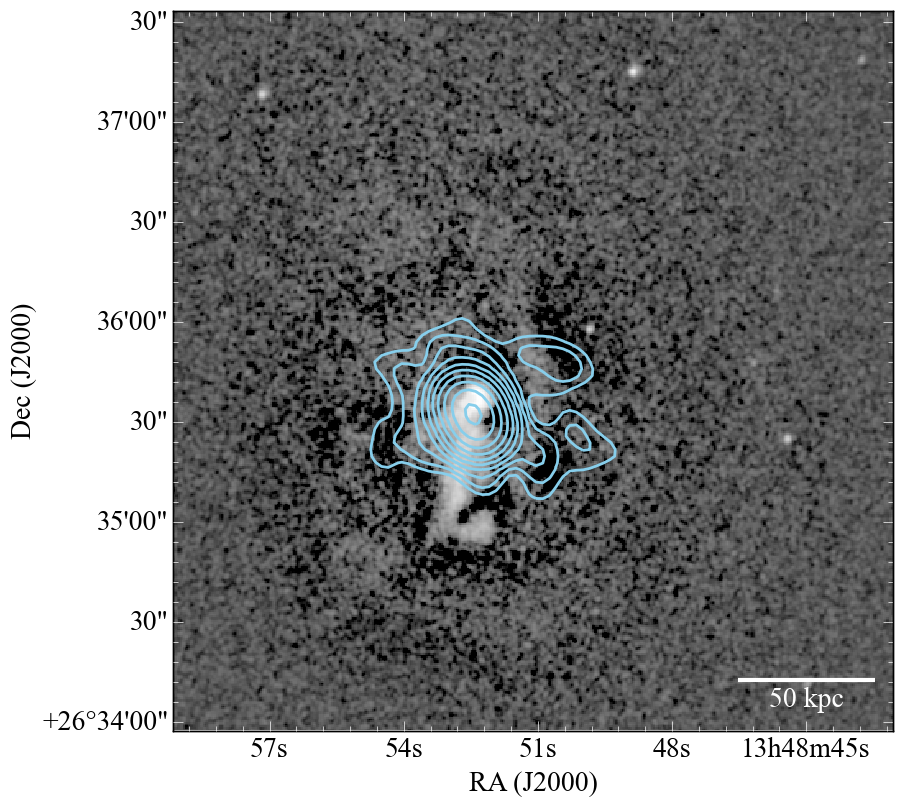} 
\includegraphics[height=2.90in]{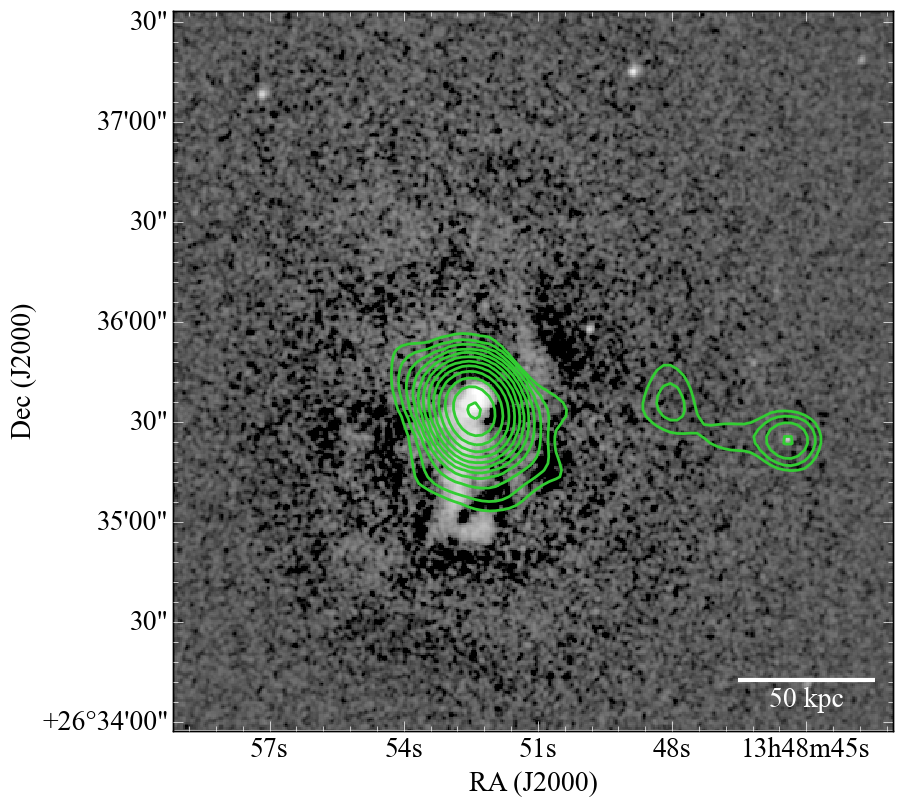}\\
\includegraphics[height=2.90in]{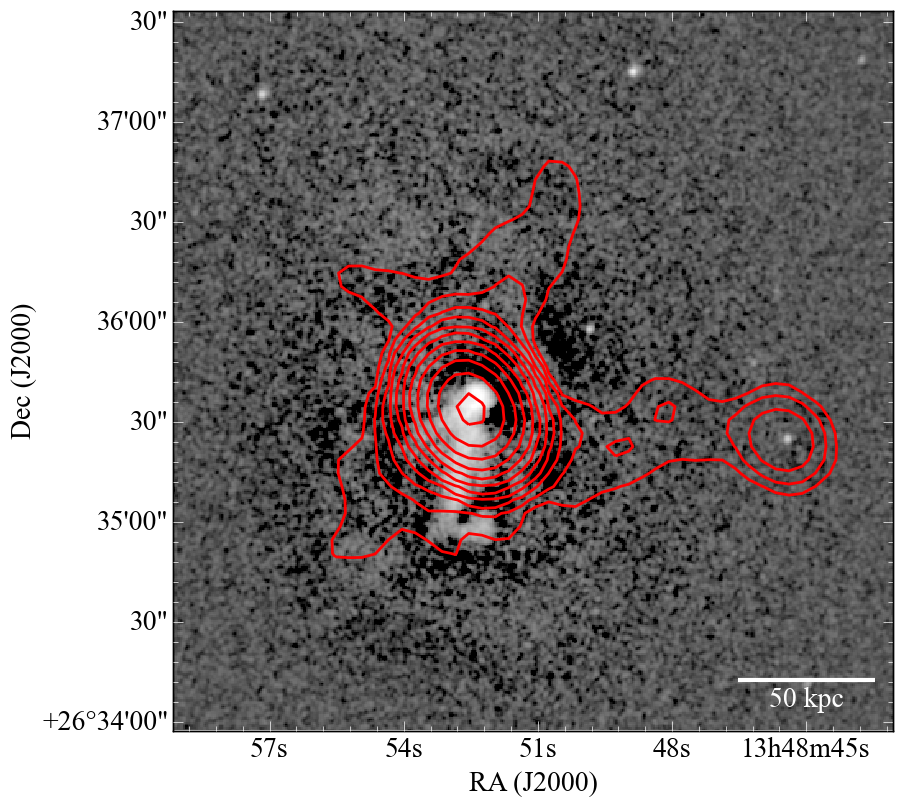} 
\includegraphics[height=2.90in]{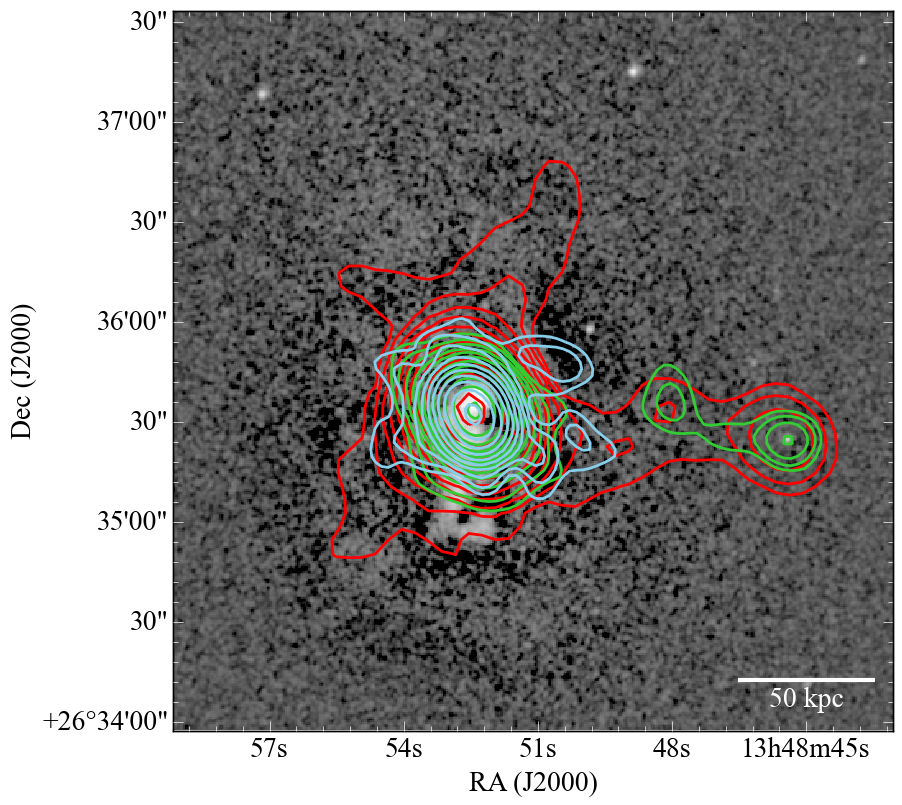}
%% 1 column
% \includegraphics[height=2.0in]{Xray_figures/A1795_Xray_res_radio_02_235MHz.png} 
% \includegraphics[height=2.0in]{Xray_figures/A1795_Xray_res_radio_03_610MHz.png}\\
% \includegraphics[height=2.0in]{Xray_figures/A1795_Xray_res_radio_04_Simona.png} 
% \includegraphics[height=2.0in]{Xray_figures/A1795_Xray_res_radio_01_all.png}
\end{center}
\caption{\small X-ray residual map of A1795 with radio contours from different radio frequencies. 
\textit{Top left:} Contours from the GMRT image at 235 MHz.
\textit{Top right:} Contours from the GMRT image at 610 MHz.
The contour levels at 610 and 235 MHz are the same as in Fig. \ref{figP2:GMRT}.
\textit{Bottom left:} Contours at 1.4 GHz derived from the VLA C-configuration image of \cite{Giacintucci2014} are shown in red. They are drawn at 5$\sigma$ $\times$ [1, 2, 4, 8, 16, 32, 64, 128, 256, 512, 1024, 2048], where the rms noise level is 0.15 mJy/beam. 
\textit{Bottom right:} Composite depicting the radio contours at 1.4 GHz C-configuration (red), 610 MHz (green), and 235 MHz (blue). 
\label{figP2:X-ray_radio_contours}}
\vspace{0.15in}
\end{figure*}

\subsection{GMRT 610 and 235 MHz}

The radio map at 610 MHz (Fig. \ref{figP2:GMRT}) shows a round, not well-resolved morphology at the core of A1795.
The 610 MHz image also reveals the tail of a nearby background source at $z_{\rm{phot}} = 0.57 \pm 0.08$ \citep{Giacintucci2014} situated to the west of the the central galaxy of A1795.
The 235 MHz  map shows a smooth morphology close to the center of A1795 with a few marginally resolved extensions toward the NE, E, and S.
The southern extension covers the western part of the X-ray tail (Fig. \ref{figP2:X-ray_radio_contours}). We study this correspondence in Sect. \ref{secP2:tail}.
The most pronounced features in the 235 MHz GMRT map are two clear extensions directed to the west of the active core.
These extensions are significant at a 5$\sigma$ level.
The radio emission of the northern extension spans up to $\sim$45 kpc from the center, thus reaching beyond the NW edge, in the region of the NW X-ray depression (Fig. \ref{figP2:X-ray_radio_contours}).
The southern feature extends up to $\sim$50 kpc and is associated with the SW X-ray depression.
We explore the correspondence between the 235 MHz radio extensions and the X-ray depressions in Sect. \ref{secP2:cavities}.\\

\subsection{VLA 1.4 GHz C-configuration}

With the help of 1.4 GHz VLA observations in C-configuration, \cite{Giacintucci2014} have studied the more extended radio structure of A1795 (Fig. \ref{figP2:X-ray_radio_contours}). They find a candidate mini-halo extending up to a radius of $\sim$100 kpc and measure a power for the mini-halo of $(7.9\pm0.5)\times10^{23}$ W Hz$^{-1}$.
An important consequence of the map of \cite{Giacintucci2014} is that the NW depression is not covered by radio emission. 
Moreover, the high-frequency radio emission seems to surround the X-ray cavity, instead of filling it. 
Thus, the inner edge of the northern cavity resembles the southern ``bay'' feature in Perseus \citep{Fabian2011}.
This is one of the factors that have made \cite{Walker2017} speculate that the NW X-ray concave enhancement is not a border of a cavity but  a bay, similar to the one observed in Perseus. 
They further show that this bay may have been created by a Kelvin--Helmholtz (KH) instability in the cold front surrounding the core of A1795. 

The radio emission that appears to surround the NW X-ray cavity consists of two features,  a radio filament toward the NW  and a bridge of emission between the core and a source to the west.
Although the NW extension in the 1.4 GHz map of \cite{Giacintucci2014} is formally detected at the 5$\sigma$ significance level, it is possibly related to calibration uncertainty since it lies along one of the stripes of the typical VLA array pattern revealed in many images and resulting from the Y-shaped arrangement of the VLA antennas.
If  this feature were interpreted as being part of the mini-halo, then we would expect it to consist of older emission and therefore be more pronounced at lower frequencies. However, this extension does not appear in the GMRT maps, which supports our suspicion that the NW extension in the 1.4 GHz map is significantly enhanced by a calibration artifact.
Furthermore, we show that the bridge of radio emission between the central galaxy and the background source to the west is likely due to chance alignment (Fig. \ref{figP2:X-ray_radio_contours}).
Our GMRT map at 610 MHz reveals that the background source has a wide tail of emission, with the plume of emission pointed toward the core of A1795. 
On the other hand, the 235 MHz image shows an extension in the radio emission of A1795 stretching toward the SW.
The simplest explanation is that the combination of these two features at lower resolution is what gives the appearance of a bridge of emission in the 1.4 GHz C-configuration map.

%%%%%%%%%%%%%%%%%%%%%%%%%%%%%%%%%%%%%%%%%%%%%
%% CAVITIES
%%%%%%%%%%%%%%%%%%%%%%%%%%%%%%%%%%%%%%%%%%%%%

\section{Sector analysis} 
\label{secP2:cavities}

\begin{figure}[t]
\begin{center}
\includegraphics[height=2.90in]{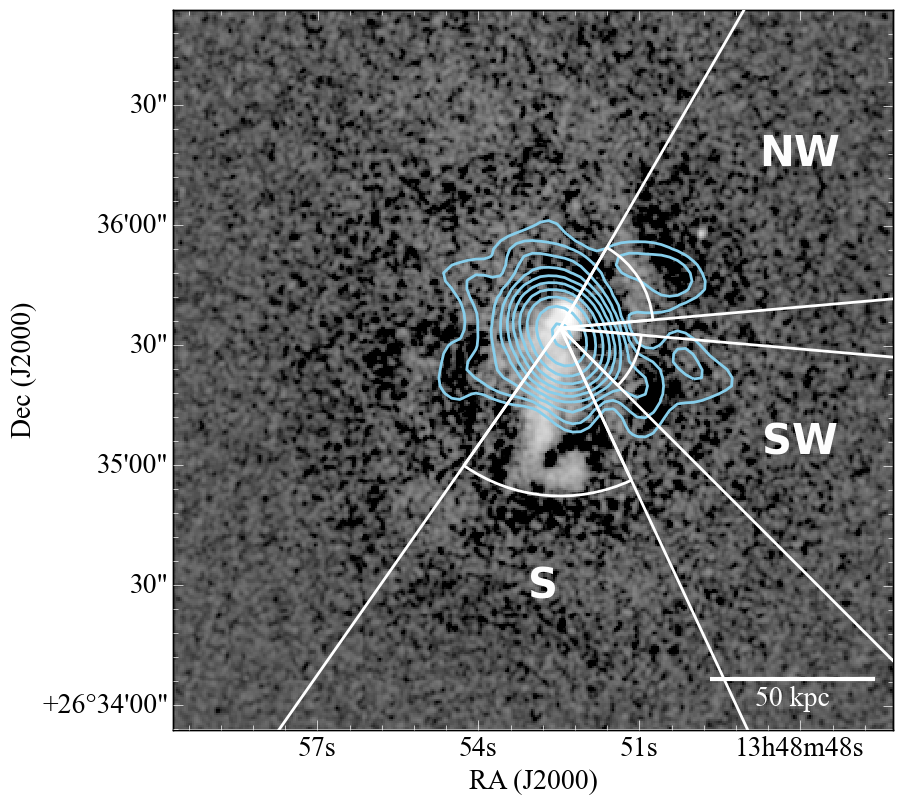}
\end{center}
\caption{\small Three sectors used for radial analysis extraction overplotted on the X-ray residual map from Fig. \ref{figP2:X-ray_residual}. The radio contours at 235 MHz are shown in blue with the same levels as in Fig. \ref{figP2:GMRT}.
Arcs in the NW, SW, and S sector show respectively the positions of the NW edge, the SW edge, and the end of the tail as determined from the X-ray residual map.
\label{figP2:wedges}}
\vspace{0.15in}
\end{figure}

\begin{figure*}[!htbp]
\begin{center}
%% 2 column
\includegraphics[height=2.75in]{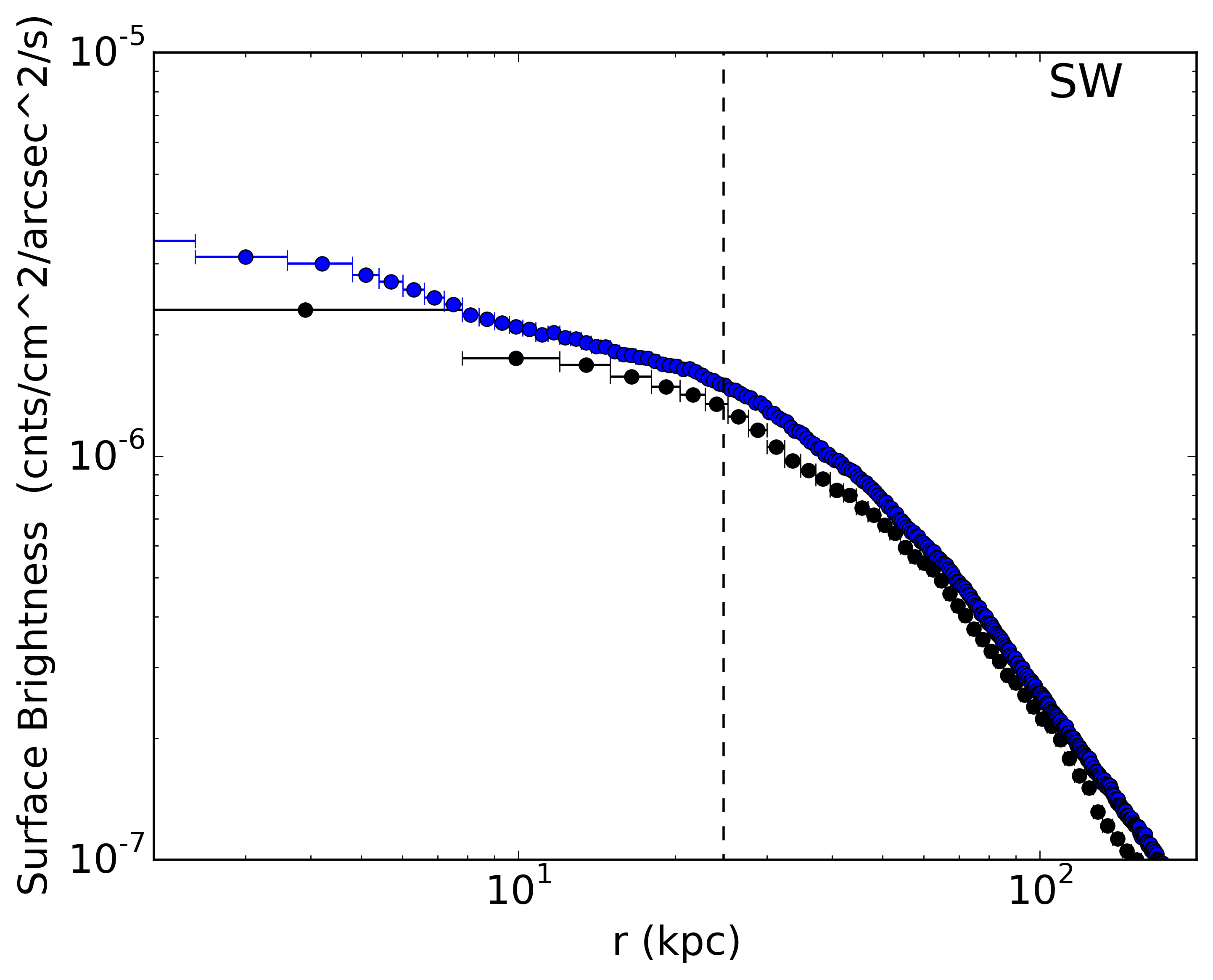} 
\includegraphics[height=2.75in]{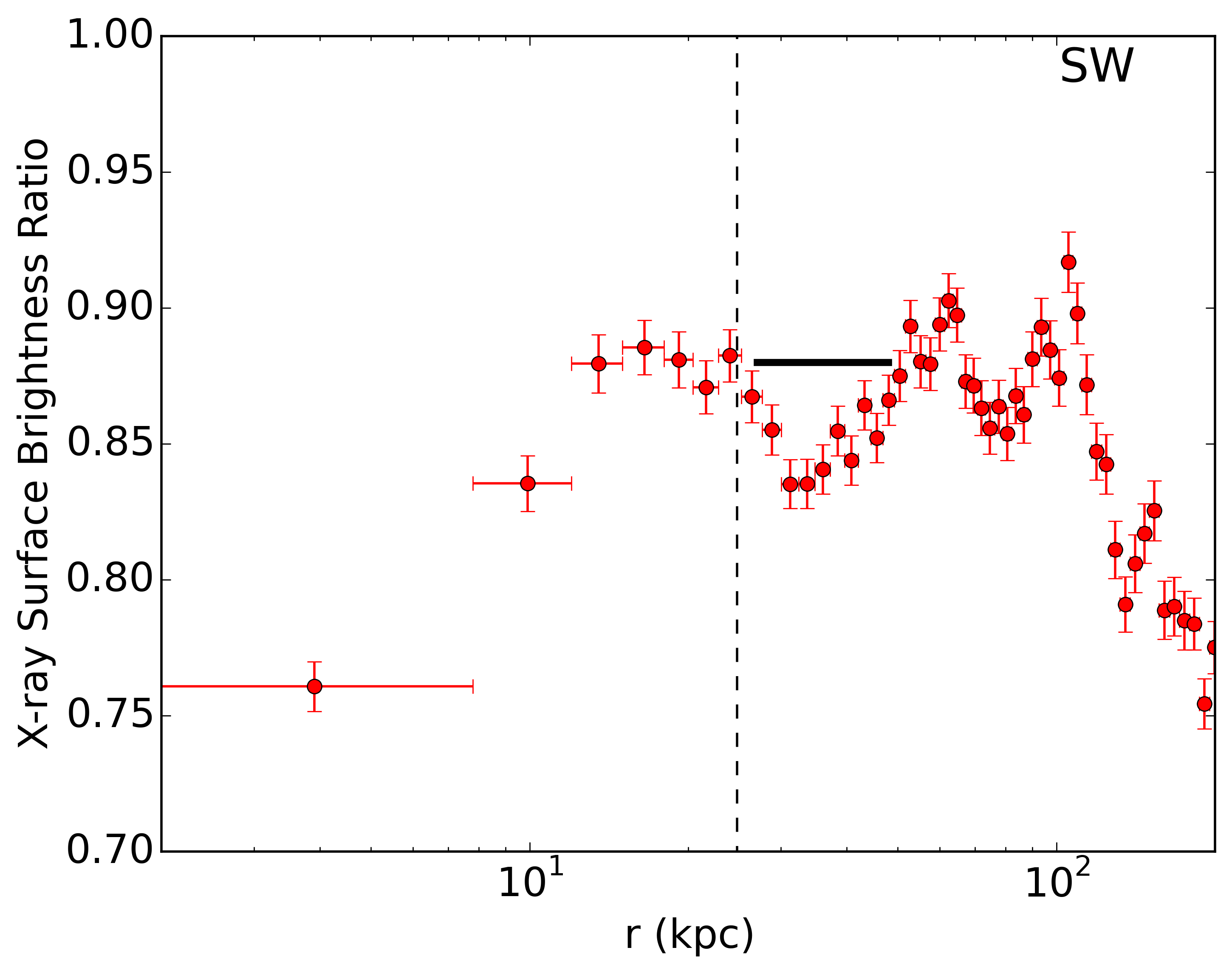} \\
\includegraphics[height=2.75in]{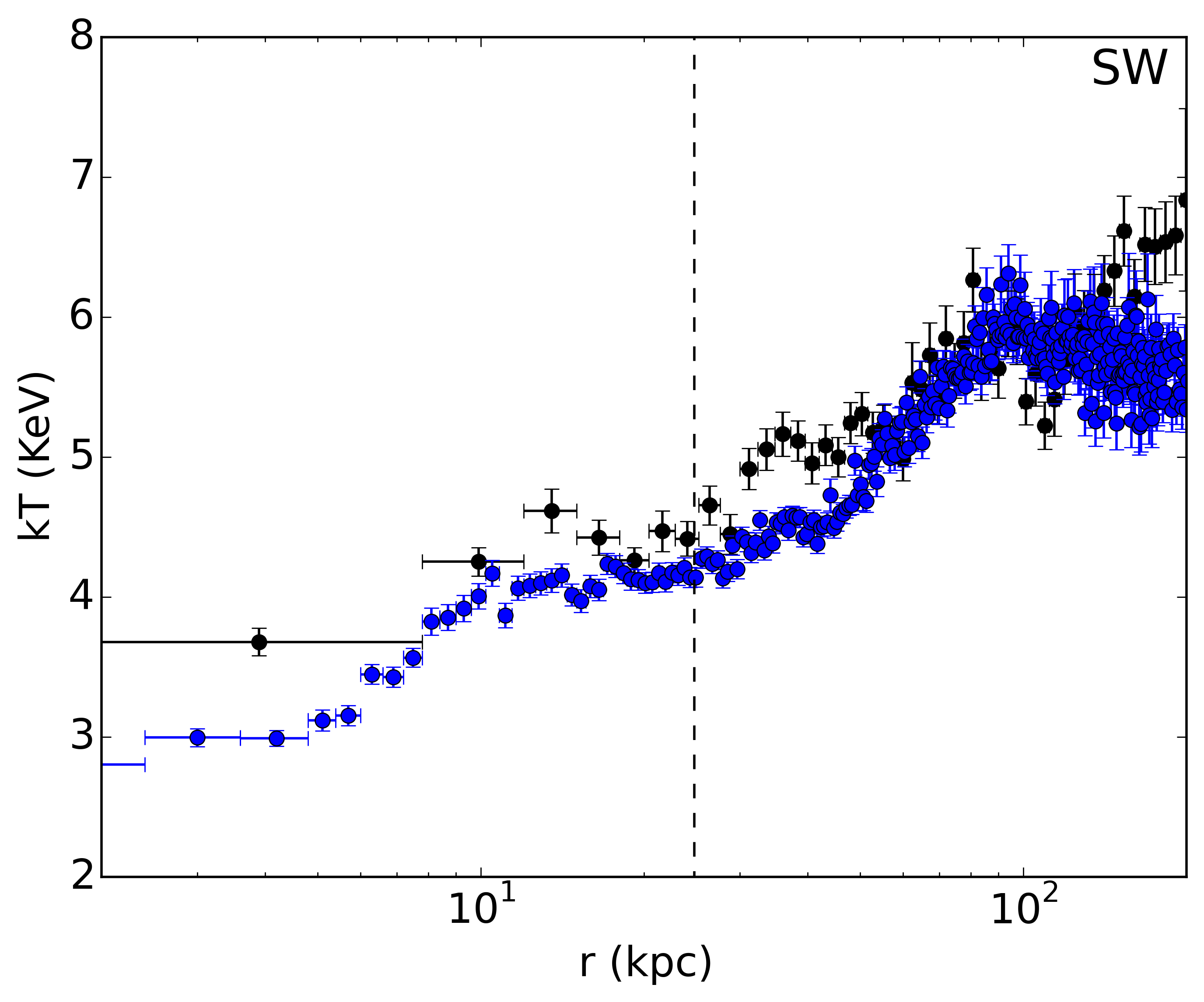} 
\includegraphics[height=2.75in]{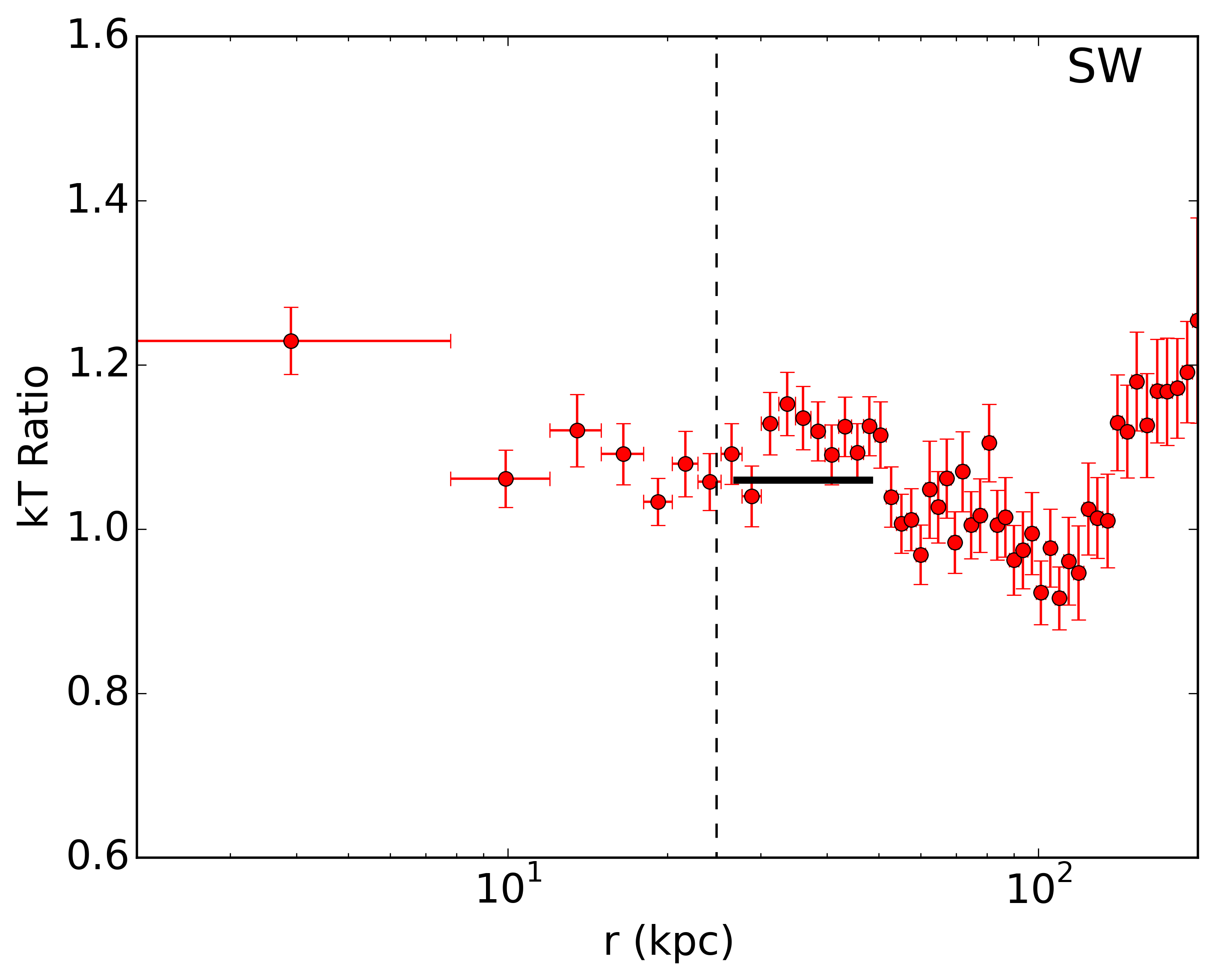}  \\
\includegraphics[height=2.75in]{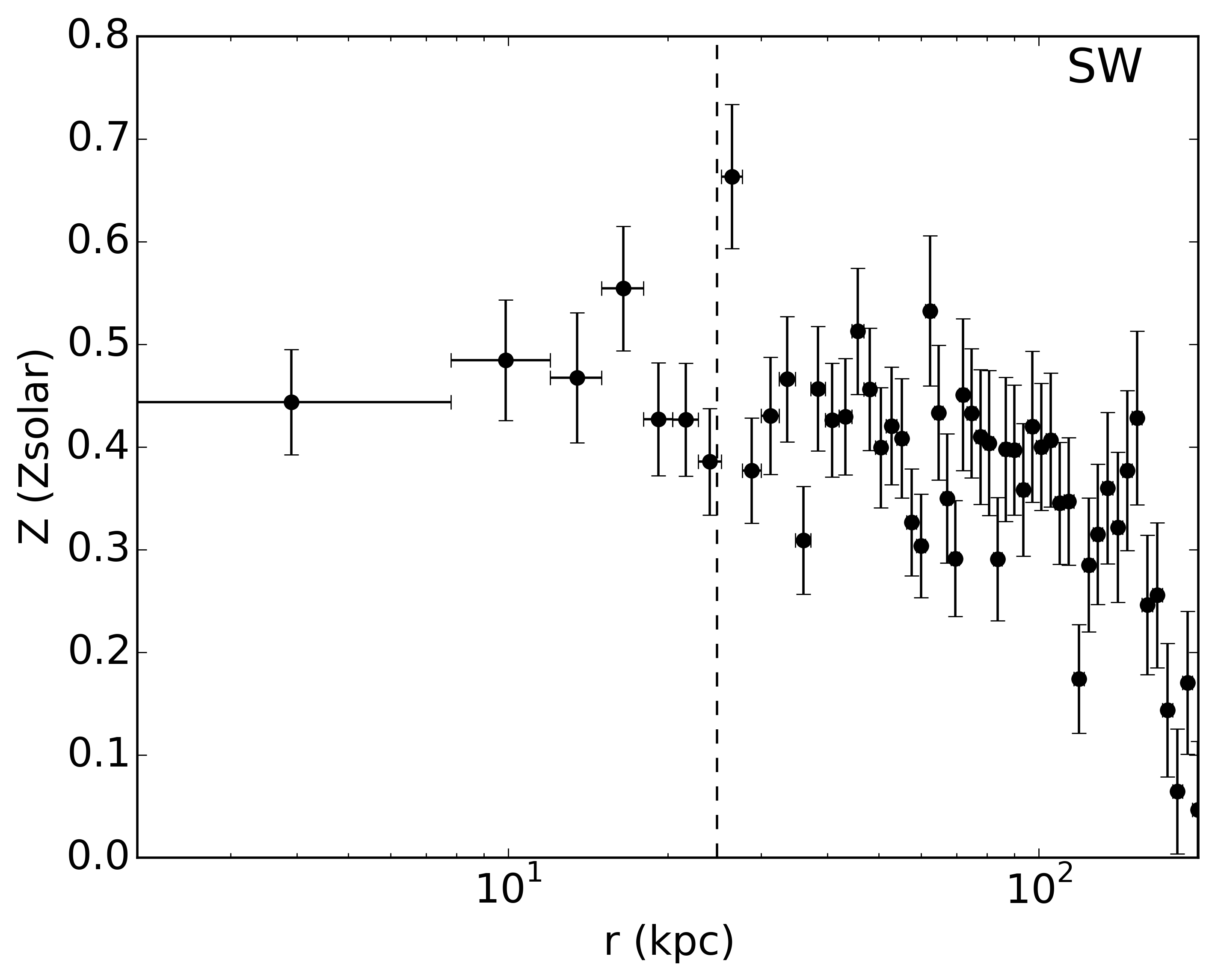} 
\includegraphics[height=2.75in]{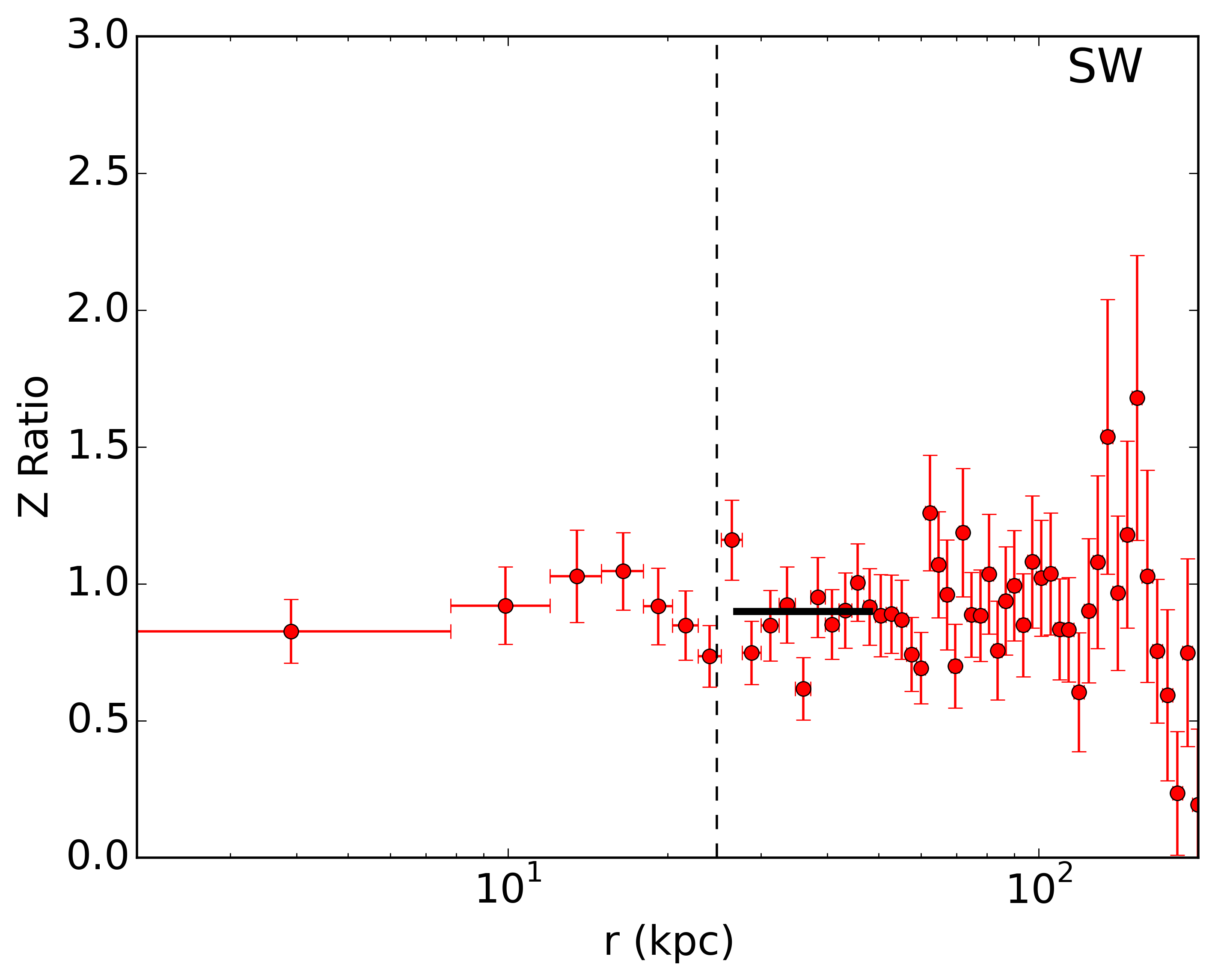}
%% 1 column
% \includegraphics[height=2.0in]{profiles_x-ray/SW_SB_kpc_aldir_big.png} 
% \includegraphics[height=2.0in]{profiles_x-ray/SW_SB_kpc_div_line.png} \\
% \includegraphics[height=2.0in]{profiles_x-ray/SW_kT_kpc_big.png} 
% \includegraphics[height=2.0in]{profiles_x-ray/SW_kT_kpc_div_line.png}  \\
% \includegraphics[height=2.0in]{profiles_x-ray/SW_Z_kpc.png} 
% \includegraphics[height=2.0in]{profiles_x-ray/SW_Z_kpc_div_line.png}
\end{center}
\caption{\small X-ray radial profiles along the SW direction.
\textit{Top left:} Surface brightness profile along the SW (black) and average surface brightness profile (blue).
\textit{Top right:} Ratio of the X-ray surface brightness profile toward the SW to the average X-ray surface brightness profile.
\textit{Middle left:} Temperature profile along the SW (black) and average temperature profile (blue).  
\textit{Middle right:} Ratio of the temperature profile toward the SW to the average temperature profile.
\textit{Bottom left:} Metallicity profile along the SW.  
\textit{Bottom right:} Ratio of the metallicity profile toward the SW to the average metallicity profile. The average metallicity profile can be seen in Fig. \ref{figP2:radial_xray_s}.
The black stripe shows the extent of the SW depression as constrained from the X-ray surface brightness ratio profile. 
The vertical blue line shows the position of the SW edge as determined from the X-ray residual map (Fig. \ref{figP2:wedges}).
\label{figP2:radial_xray_sw}}
\vspace{0.15in}
\end{figure*}

Based on the discussion on the X-ray and radio morphology in the previous two sections we  defined three wedge-shaped regions of interest centered on the BCG.
To support our analysis, we  extracted radial profiles confining regions along these major directions (Fig. \ref{figP2:wedges}).  
The NW profile follows the NW cavity and the northern radio extension at 235~MHz, while the SW wedge stretches along the SW cavity and the southern radio extension. 
The cold tail is covered by the southern profile.
In addition we  extracted an averaged radial profile in all directions.
We present surface brightness, temperature, and metallicity profiles extracted from the X-ray data (Fig. \ref{figP2:radial_xray_nw} and \ref{figP2:radial_xray_sw}), as well as radio surface brightness profiles at 235~MHz (Fig. \ref{figP2:radio_profiles}).

To construct the X-ray profiles the size of the bins in the radial direction is adapted so that each bin reaches a SN of 100.
The radial extent of each bin in the radio profiles corresponds to the FWHM of the synthesized beam of the 235~MHz image (shown in Fig. \ref{figP2:GMRT}).
The associated error is given by $\sigma_{n}\times\sqrt{N_b}$, where $\sigma_{n}$ is the measured rms noise in the image and $N_b$ is the number of beams sampled in each bin.

In this section we discuss the profiles in the NW and SW directions.
The profile in the S is analyzed in Sect. \ref{secP2:tail} where we study the properties of the cold tail.

\subsection{NW sector}

\begin{figure*}[!htbp]
\begin{center}
%% 2 column
\includegraphics[height=2.75in]{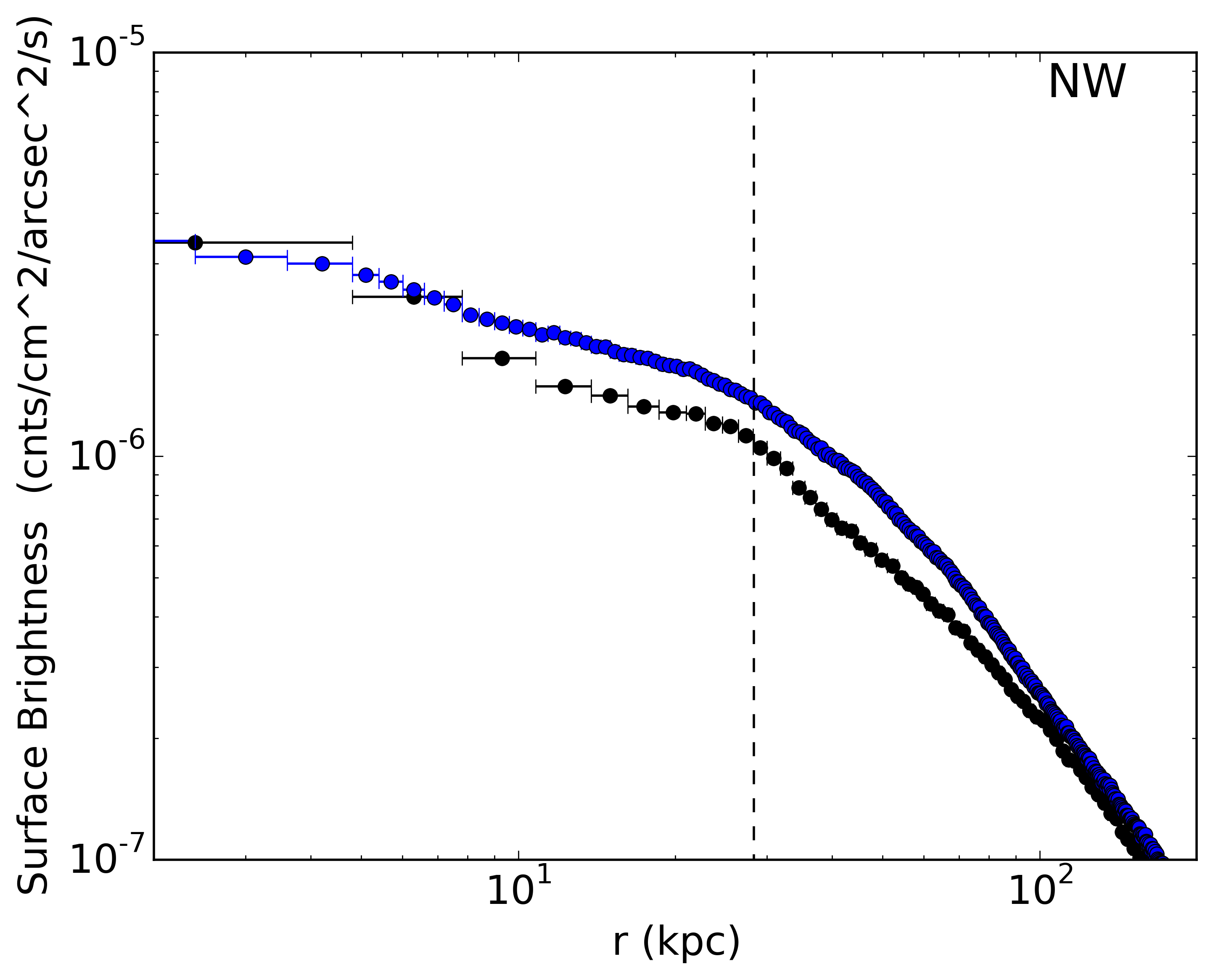} 
\includegraphics[height=2.75in]{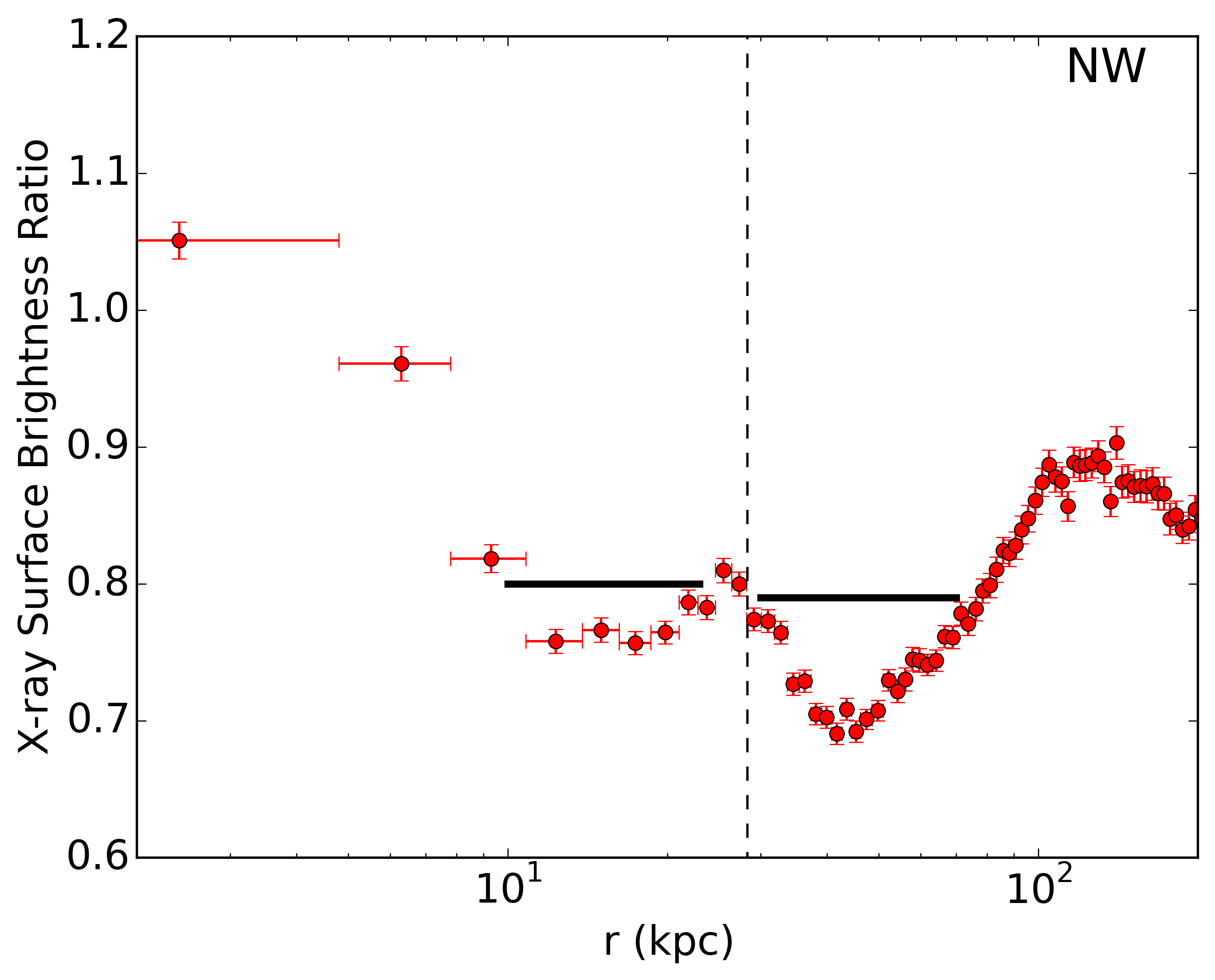} \\
\includegraphics[height=2.75in]{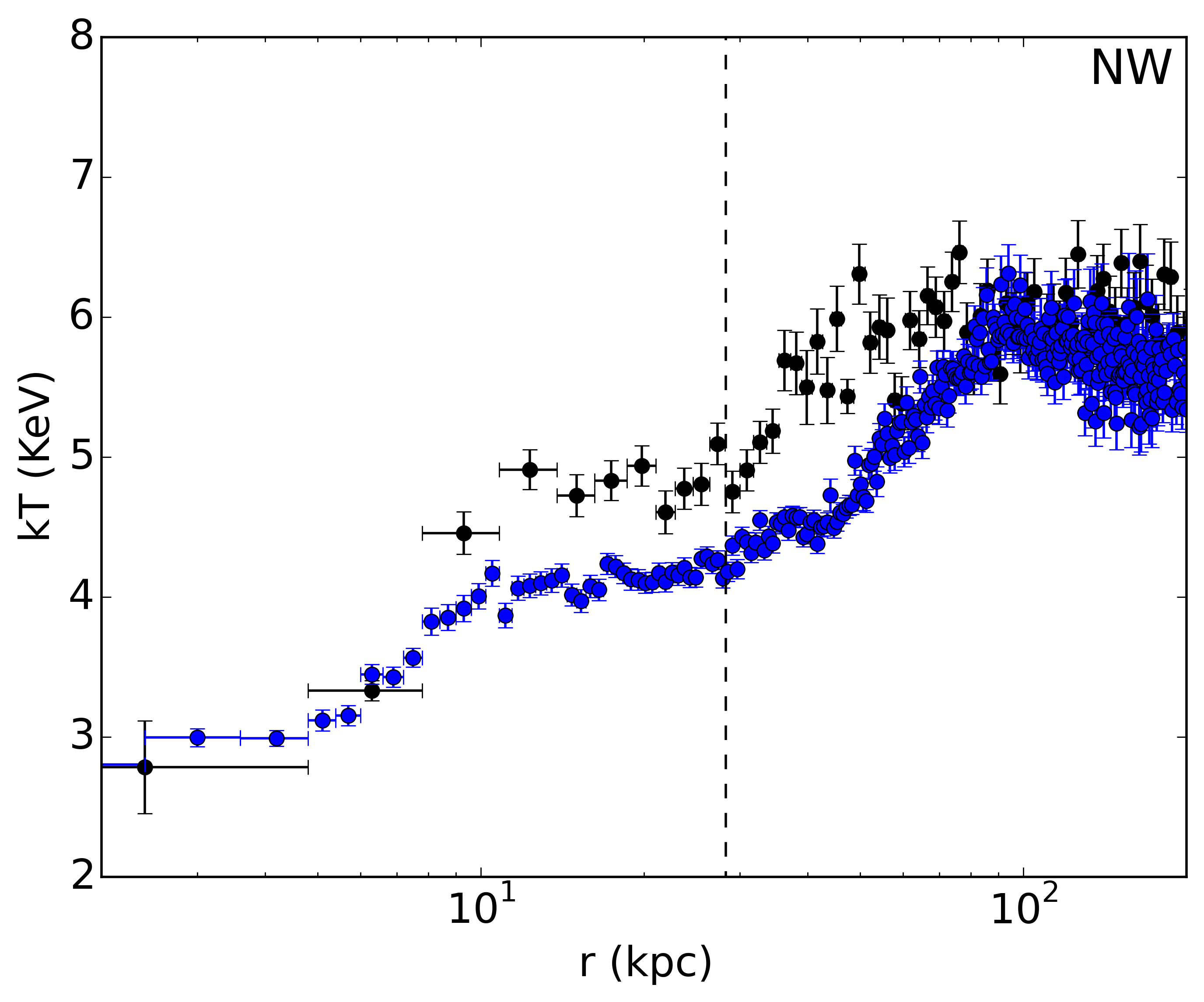} 
\includegraphics[height=2.75in]{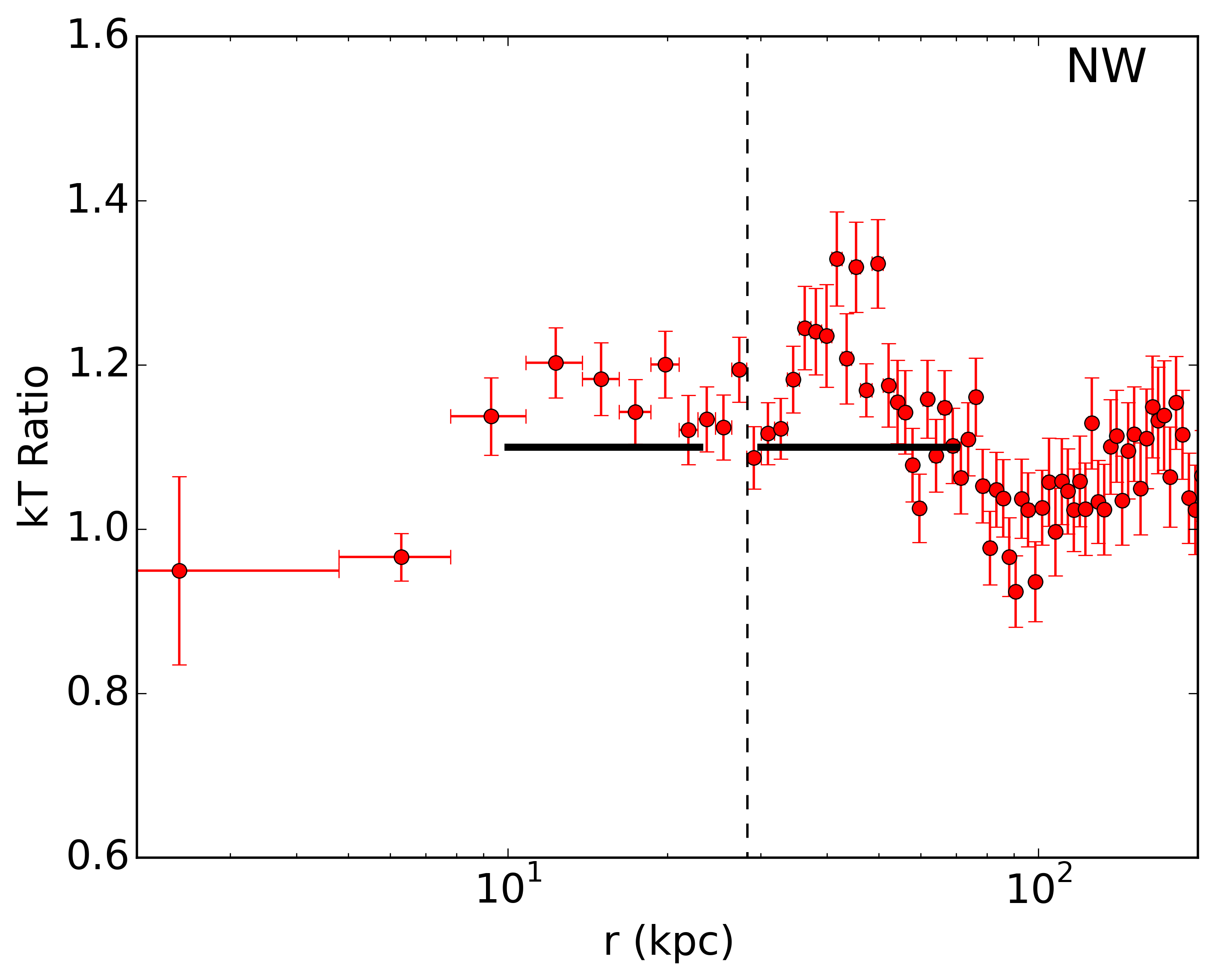} \\
\includegraphics[height=2.75in]{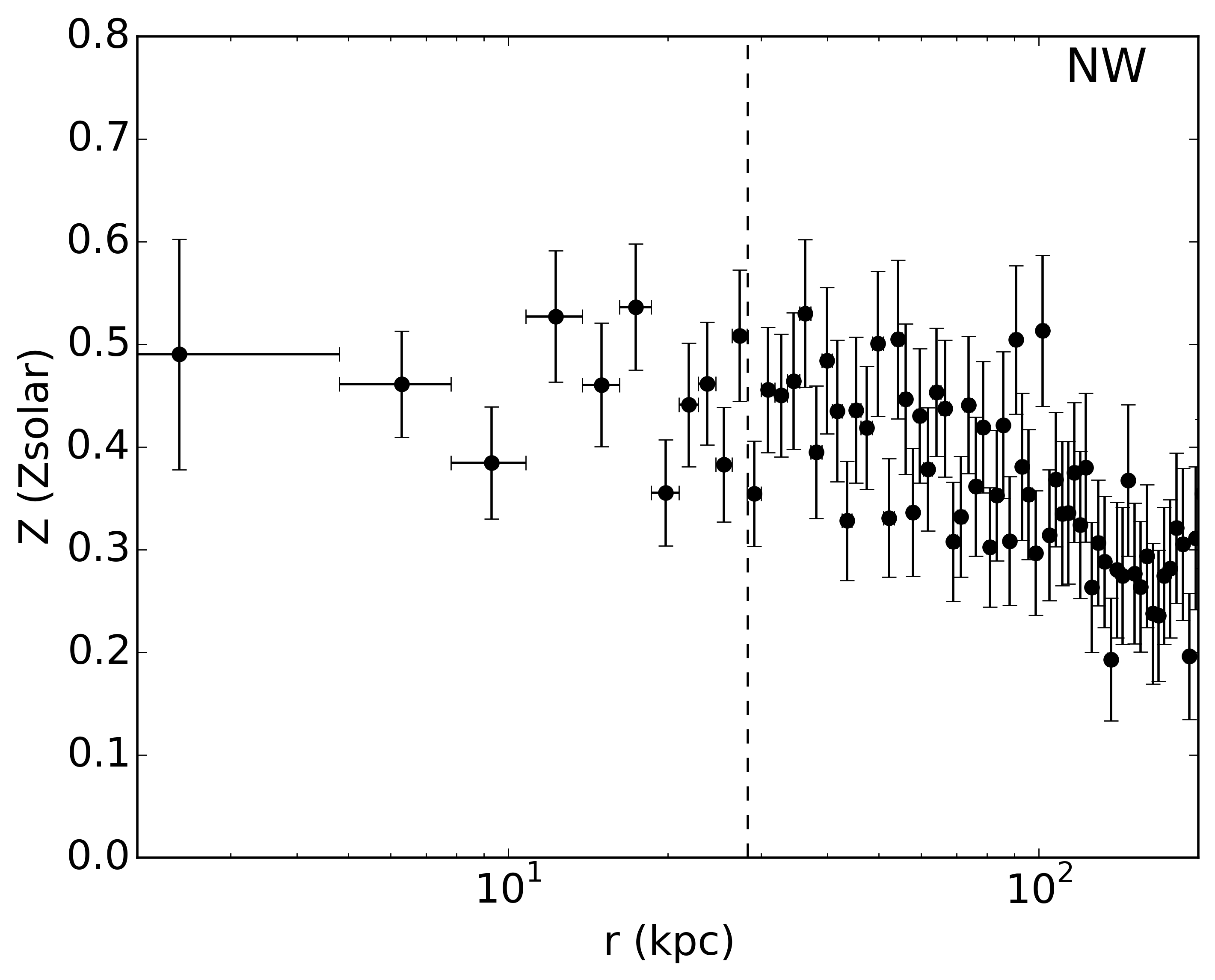} 
\includegraphics[height=2.75in]{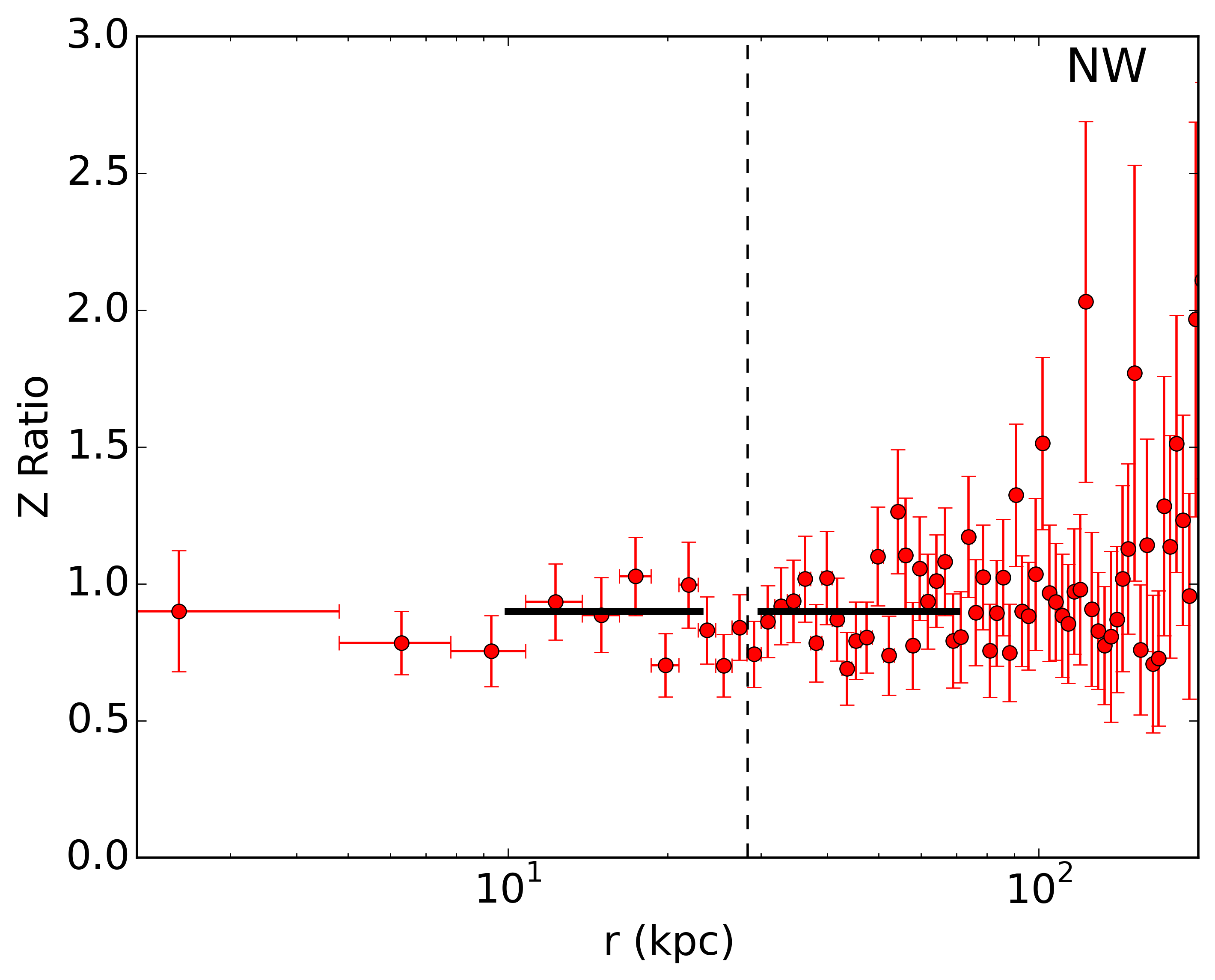} 
%% 1 column
% \includegraphics[height=2.0in]{profiles_x-ray/NW_SB_kpc_aldir_big.png} 
% \includegraphics[height=2.0in]{profiles_x-ray/NW_SB_kpc_div_line.png} \\
% \includegraphics[height=2.0in]{profiles_x-ray/NW_kT_kpc_big.png} 
% \includegraphics[height=2.0in]{profiles_x-ray/NW_kT_kpc_div_line.png} \\
% \includegraphics[height=2.0in]{profiles_x-ray/NW_Z_kpc.png} 
% \includegraphics[height=2.0in]{profiles_x-ray/NW_Z_kpc_div_line.png} 
\end{center}
\caption{\small X-ray radial profiles along the NW direction.
\textit{Top left:} Surface brightness profile along the NW (black) and average surface brightness profile (blue).
\textit{Top right:} Ratio of the X-ray surface brightness profile toward the NW to the average X-ray surface brightness profile.
\textit{Middle left:} Temperature profile along the NW (black) and average temperature profile (blue).  
\textit{Middle right:} Ratio of the temperature profile toward the NW to the average temperature profile.
\textit{Bottom left:} Metallicity profile along the NW.   
\textit{Bottom right:} Ratio of the metallicity profile toward the NW to the average metallicity profile. The average metallicity profile can be seen in Fig. \ref{figP2:radial_xray_s}.
The black stripes show the extent of the depressions as constrained from the X-ray surface brightness ratio profile. 
The vertical dashed line indicates the position of the NW edge as determined from the X-ray residual map (Fig. \ref{figP2:wedges}).
\label{figP2:radial_xray_nw}}
\vspace{0.15in}
\end{figure*}

Figure \ref{figP2:radial_xray_nw} shows the surface brightness profile along the NW sector. In the top left  panel the NW profile is compared with the average profile.
This comparison suggests two regions where the NW profile is much lower than the average value.
To highlight the difference between the two profiles we have plotted the ratio of the NW to the average profile in the top right  panel of Fig. \ref{figP2:radial_xray_nw}. 
This ratio profile shows two pronounced dips (cavity-like structures) at the two sides of the NW edge. 
These regions correspond to the NW and the arc depressions identified in the residual map.

The ratio of the temperature profiles (Fig. \ref{figP2:radial_xray_nw}) shows two temperature enhancements corresponding to the two dips in the X-ray surface brightness.
While the surface brightness dips might be suggestive of shock fronts or cold fronts due to sloshing, the temperature profile does not support these possibilities.
The temperature profiles do not exhibit any sharp jumps or drops, but rather a smooth increase in temperature peaking at the mid-length of the surface brightness depressions. This temperature behavior is more consistent with a cavity structure produced by the AGN injecting hot plasma within the bubbles.

Based on the ratio of the X-ray surface brightness profiles, the outer depression is situated 44 kpc from the center and has a radius of 10 kpc, assuming spherical symmetry.
This corresponds to an age of 38 Myr, based on the sound speed in this region.
Assuming the cavity is filled  with relativistic plasma, its total energy is $E = 4pV = 4 \times 10^{59} \rm{erg}$
and the power of the outburst is estimated to be $7\times10^{44}$erg/s.

While the NW cavity is one of the most distant cavities observed, the measured energy and power are close to the average values known from other systems.
We note that the total energy we compute is an order or magnitude lower than the value presented by \cite{Walker2014}. 
This follows from the  size they infer from the curvature of the inner rim of the cavity, which is almost two times larger. The difference in size of the cavity also contributes to the longer sound crossing time (41 Myr) they measure.

The arc depression has a radius of 7 kpc. 
It is located at a distance 16 kpc from the center, which corresponds to an age of 14 Myr.
Since this cavity is clearly non-spherical, we estimate its volume based on a part of a torus with a length of 23 kpc derived from the residual map.
We compute a total energy of $5 \times 10^{58} \rm{erg}$ and infer a power of the AGN outburst of $9\times10^{43}$erg/s. If the NW depression and the arc depression indeed resulted from the AGN activity, then the outbursts that created them have almost an order of magnitude difference in power.

\begin{figure}[!htbp]
\begin{center}
\includegraphics[height=2.6in]{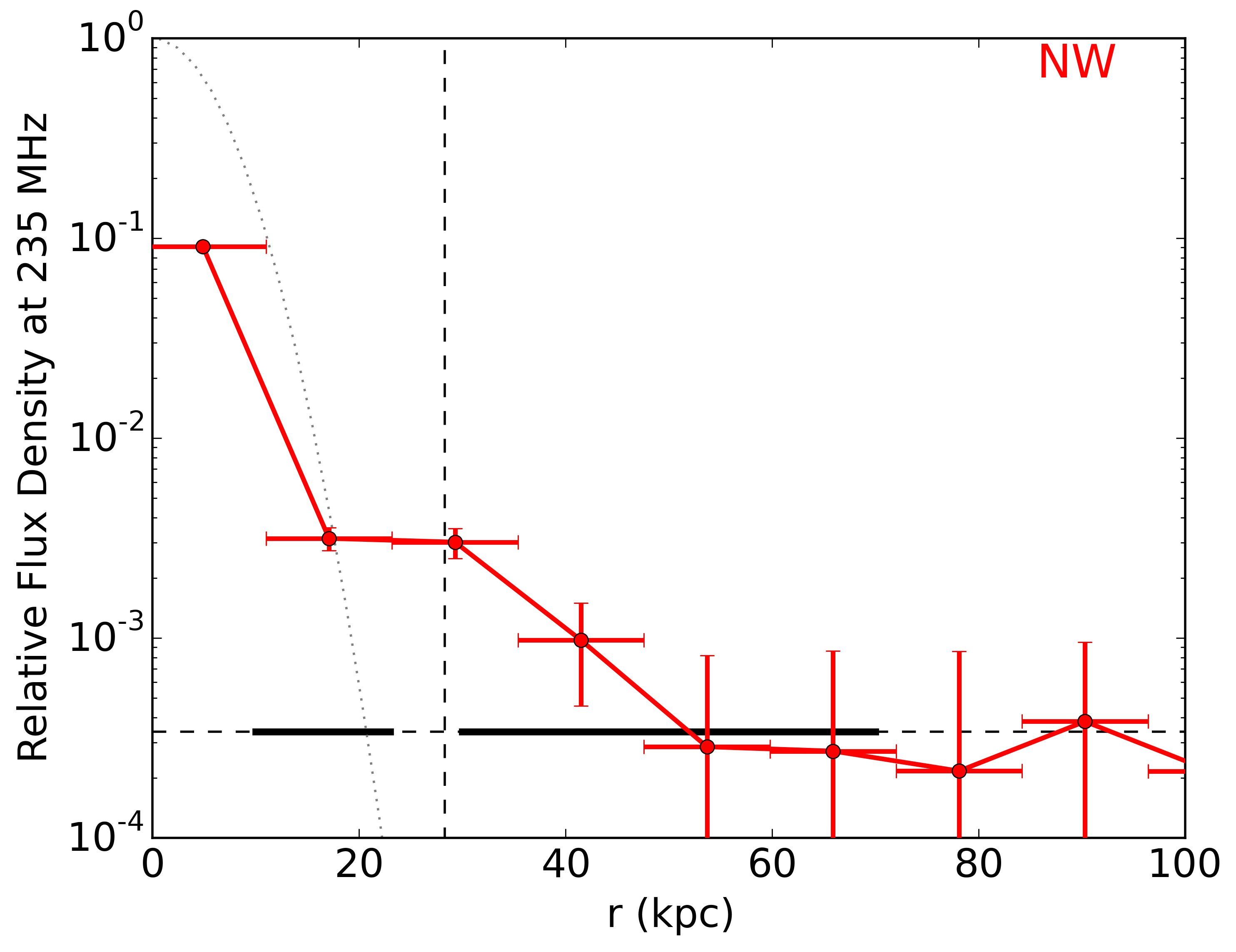}\\
\includegraphics[height=2.6in]{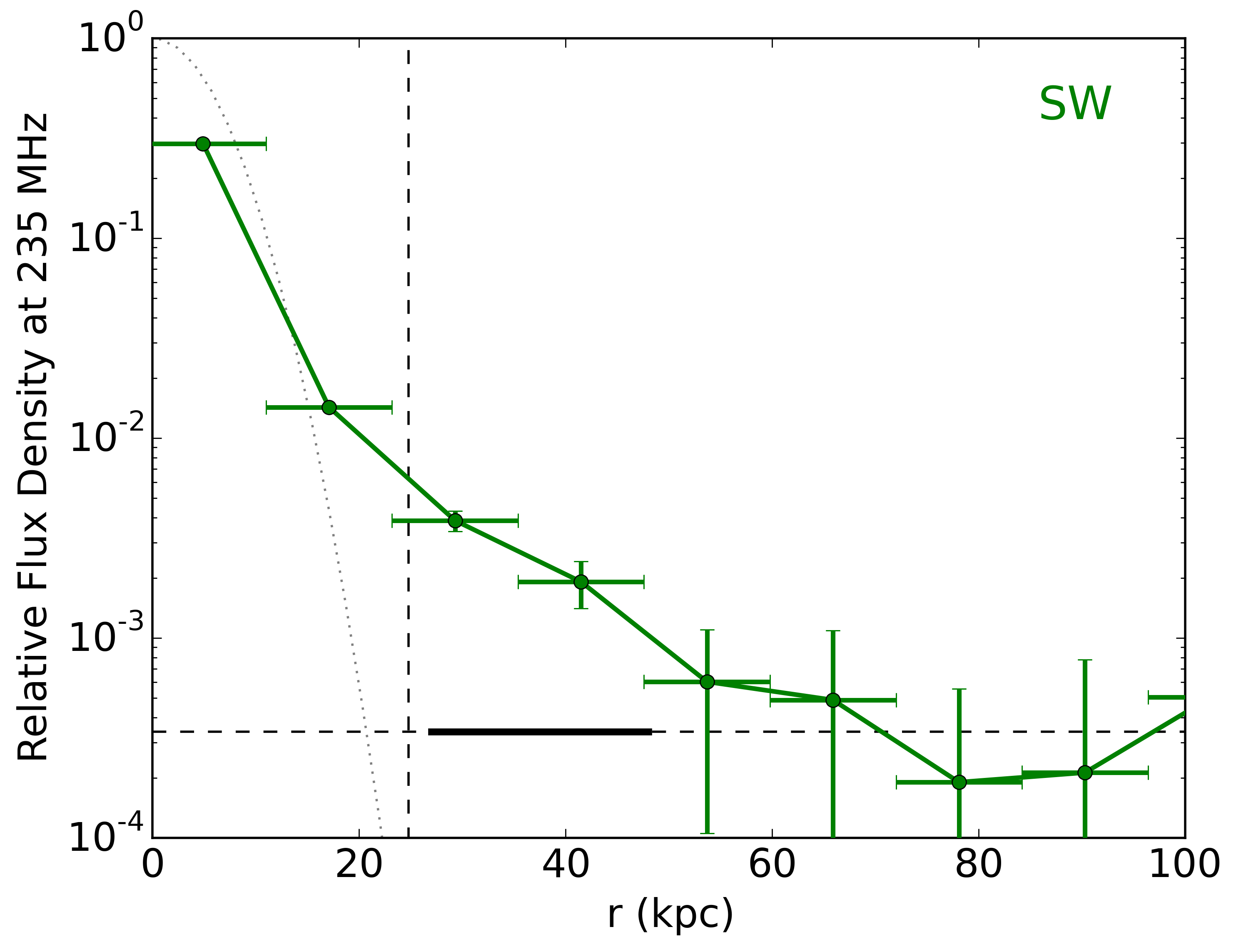}\\
\includegraphics[height=2.6in]{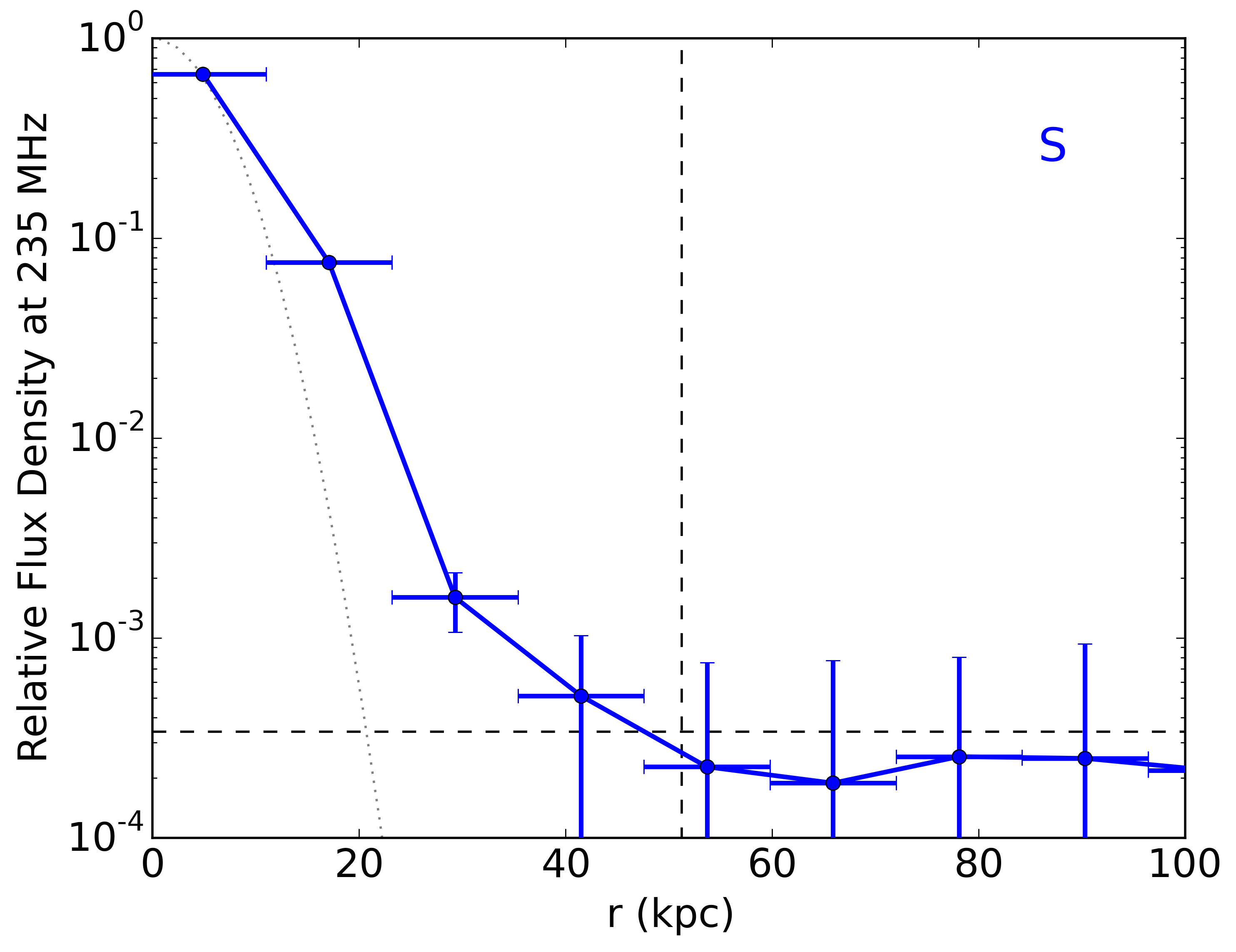}
\end{center}
\caption{\small Radial radio profiles at 235 MHz in the NW (\textit{top}), SW (\textit{middle}), and S (\textit{bottom}) directions. 
The vertical lines give the positions of the NW edge, SW edge, and the end of the S tail, as determined from the X-ray residual map (Fig. \ref{figP2:wedges}).
The horizontal dashed line indicates the rms noise level in the image. 
The black stripes show the extent of the depressions as constrained from the X-ray surface brightness ratio profiles (Fig. \ref{figP2:radial_xray_nw} and \ref{figP2:radial_xray_sw}). 
A Gaussian characteristic of the beam is shown with a gray dotted line to emphasize that the observed extended radio structures are significant in comparison with the beam size.
The flux density is presented in relative units with respect to the peak flux measured at the center of A1795. 
\label{figP2:radio_profiles}}
\vspace{0.15in}
\end{figure}

The top panel of Fig. \ref{figP2:radio_profiles} presents the radial profile extracted from the 235~MHz map along the NW radio extension.
The plotted profile shows clearly extended radio emission in the NW direction. 
The statistically significant emission at 235 MHz reaches beyond the NW edge, up to $\sim$40~kpc from the center, and covers the inner part of the NW cavity. 
As already seen from the maps in Fig. \ref{figP2:X-ray_radio_contours}, the arc cavity is fully covered by low-frequency radio emission.
The fact that we observe radio emission corresponding to the NW and the arc depressions provides another piece of strong evidence that the observed X-ray depressions are the result of previous AGN activity.
We note that it is hard to draw a definitive conclusion about the origin of the arc cavity because is shows an unusual shape and in general the ICM morphology is very complex in this direction. 
However, the radio source is clearly disturbing and distorting the medium in that direction which is consistent with the cavity interpretation for the arc depression.

\subsection{SW sector}

The SW profiles shown in Fig. \ref{figP2:radial_xray_sw} allow us to identify two regions of decreased X-ray surface brightness.
They are both located beyond the SW edge, at distances of 33 and 78 kpc from the central SMBH.
Although the SW sector passes over the arc cavity (Fig. \ref{figP2:wedges}), it is difficult to trace this cavity in the surface brightness profile.
This is likely due to the insufficient number of bins within the inner 10 kpc from the SMBH which would help us discern the inner edge of the cavity.

The inner depression distinguished in our surface brightness profile corresponds to the SW depression identified in Sect. \ref{secP2:xray_morphology}. 
We observe a significant temperature enhancement corresponding to this depression (Fig. \ref{figP2:radial_xray_sw}) which supports its classification as a cavity.
The observed temperature behavior suggests that the observed depression contains hotter plasma presumably injected by the AGN. 
An alternative explanation is that the temperature appears to rise within the cavities simply because the cavity has displaced cooler X-ray gas, thus hotter material dominates along the line of sight.
We compute an age of 30 Myr for the SW cavity, assuming propagation with the local sound speed.
The total energy of the cavity is $7 \times 10^{58} \rm{erg}$, while the cavity power is $1\times10^{44}$ erg/s.

The radio profile along the SW sector resembles the profile in the NW direction (Fig. \ref{figP2:radio_profiles}).
While the bulk of the emission is clearly restricted by the edges in both the NW and SW, the radio plasma is not fully contained by them.
Similar to the picture observed toward the NW, the SW radio profile exhibits an extension spanning up to $\sim$50~kpc from the center.
The radio plasma toward the SW appears more extended than in the NW and it fully fills in the SW cavity. 
This configuration additionally confirms that this cavity is a product of a past AGN activity episode. 

Despite our interpretation of the cavities along the NW and SW directions as the products of past AGN activity, we note that the measured abundance profiles are relatively flat over the central 100 kpc and show no evidence of significant extensions of enhanced metallicity along these directions relative to the average profile (Figs. \ref{figP2:radial_xray_nw} and \ref{figP2:radial_xray_sw}). 
At first glance, this result would seem to contradict the results of \cite{Kirkpatrick2015} who have found evidence of a metallicity extension along the axis of the radio jet in the core of A1795. The ``on-jet'' regions defined in \cite{Kirkpatrick2015}, however, only overlap with our S sector containing the S cold tail. For their ``off-jet'' direction, defined to be orthogonal to the on-jet direction and more closely aligned with our NW and SW sectors, they also found the region of enhanced metallicity in the core to be less extended. If the proposed NW and SW cavities are indeed relics of past AGN activity, they do not seem to have lifted metal-enriched gas to larger radii to the extent seen by \cite{Kirkpatrick2015} in the N and S directions.

The outer dip in our surface brightness radial profile (Fig. \ref{figP2:radial_xray_sw}) seems to be part of a large-scale spiral structure of low intensity surrounding the core.
It is evident from Fig. \ref{figP2:wedges} that only a small part of this spiral feature falls within the SW sector.
The elongated structure, which seems to stretch from the end of the tail up to the northern end of the SW sector, does not morphologically resemble a cavity, but rather suggests a shock or a cold front.
Moreover, we do not distinguish a temperature bump associated with the outer surface brightness dip.
We only find a drop in temperature, which is more consistent with a cold front.
Based on this we conclude that the outer depression in the X-ray surface brightness profile does not correspond to an AGN outflow cavity.

\subsection{Spectral index profiles}

\begin{figure}[t]
\begin{center}
\includegraphics[height=2.60in]{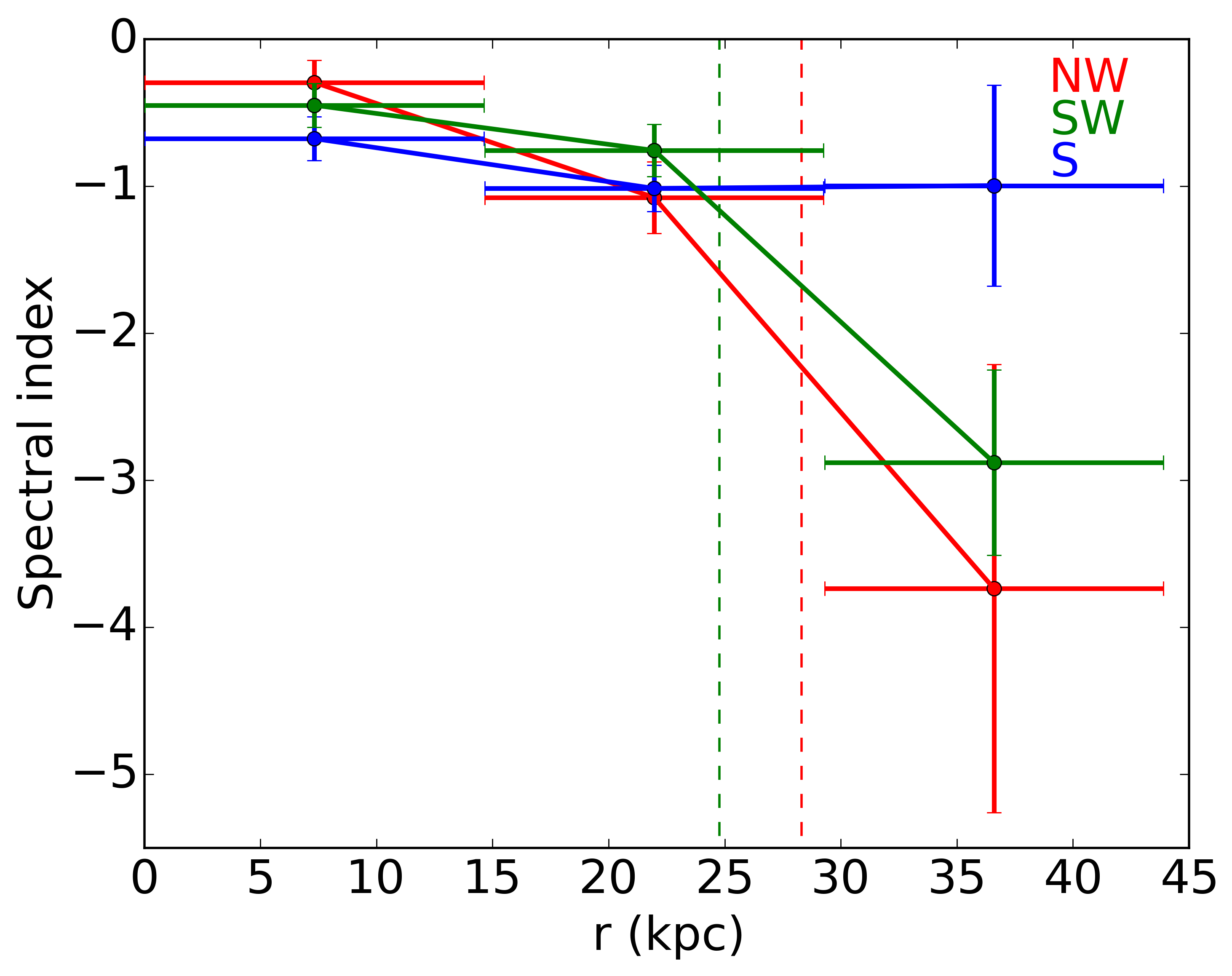}
\end{center}
\caption{\small Spectral index between 610 MHz and 235 MHz. The vertical lines show the positions of the NW (red) and the SW (green) edges.
In addition to the statistical errors, the errors of the spectral index include 10\% uncertainty in the flux scale of the two frequencies used. 
\label{figP2:spectral_index_profile}}
\vspace{0.15in}
\end{figure}

After matching the resolutions of the images at 235 and 610~MHz, we have extracted radial profiles from the two frequencies and used them to reconstruct a spectral index profile along the three chosen directions (Fig. \ref{figP2:spectral_index_profile}).
The obtained spectral index profiles reveal very different behavior of the plasma in the NW and SW compared to the S.
The spectral index of the radio emission between 235 and 610~MHz remains relatively flat in the S direction. It changes from $-0.7$ close to the center to $-1.0$ at the tail.
Similarly, the spectral index in the NW and SW stays flatter than $-1.0$ inside the NW and SW edges.
Outside the edges, however, the emission in these two directions appears very aged  exhibiting a spectral index lower than $-2.9$.
The spectral index profiles in radio are consistent with the interpretation that extensions observed in the GMRT map at 235~MHz consist of aged plasma, produced from an older outburst.
The similar spectral properties of the two extensions suggest that they have similar ages and are thus likely produced by the same AGN outburst.

%%%%%%%%%%%%%%%%%%%%%%%%%%%%%%%%%%%%%%%%%%%%%
%% TAIL
%%%%%%%%%%%%%%%%%%%%%%%%%%%%%%%%%%%%%%%%%%%%%

\section{Cold tail}
\label{secP2:tail}

\begin{figure*}[!htbp]
\begin{center}
%% 2 column
\includegraphics[height=2.75in]{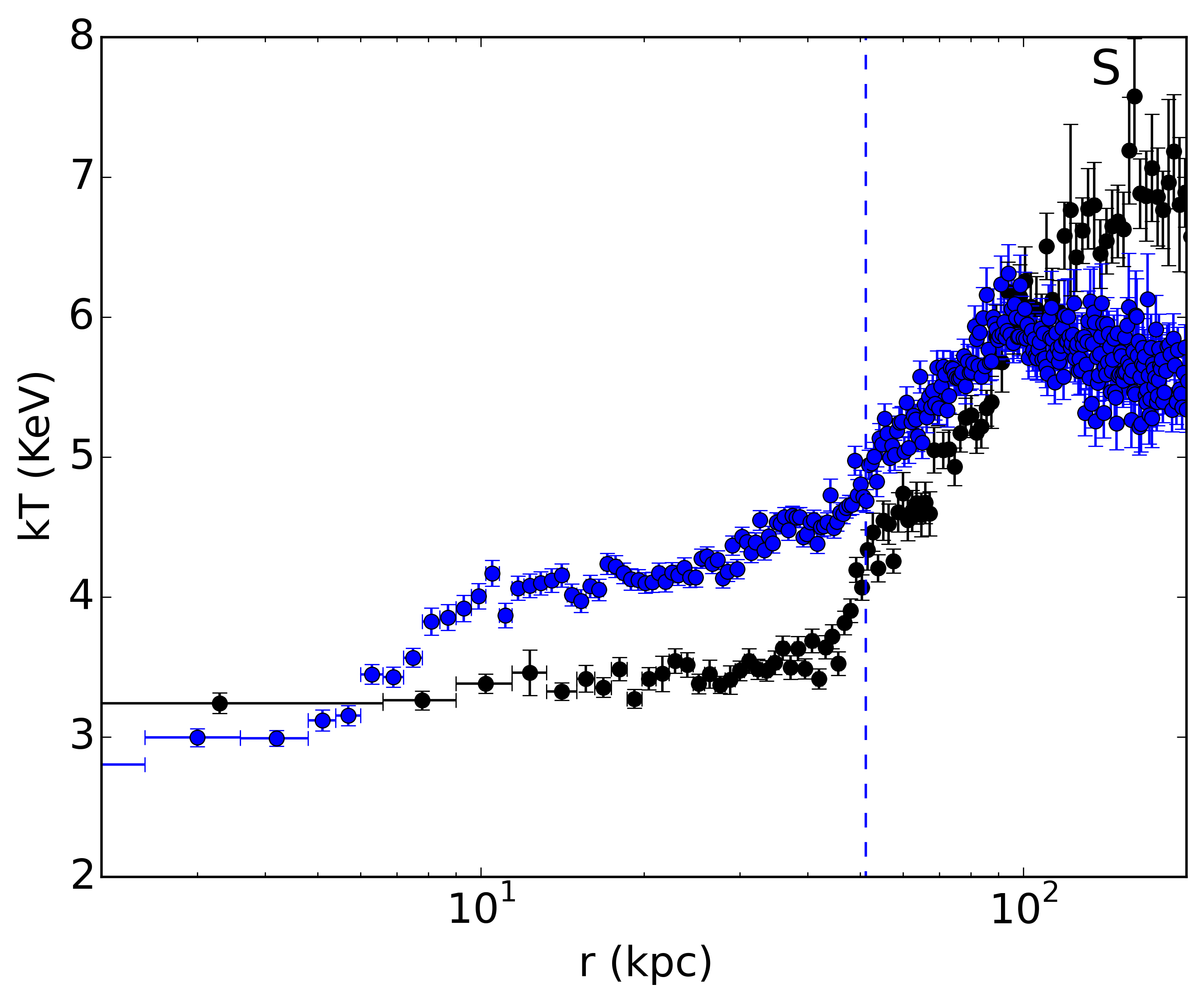} 
\includegraphics[height=2.75in]{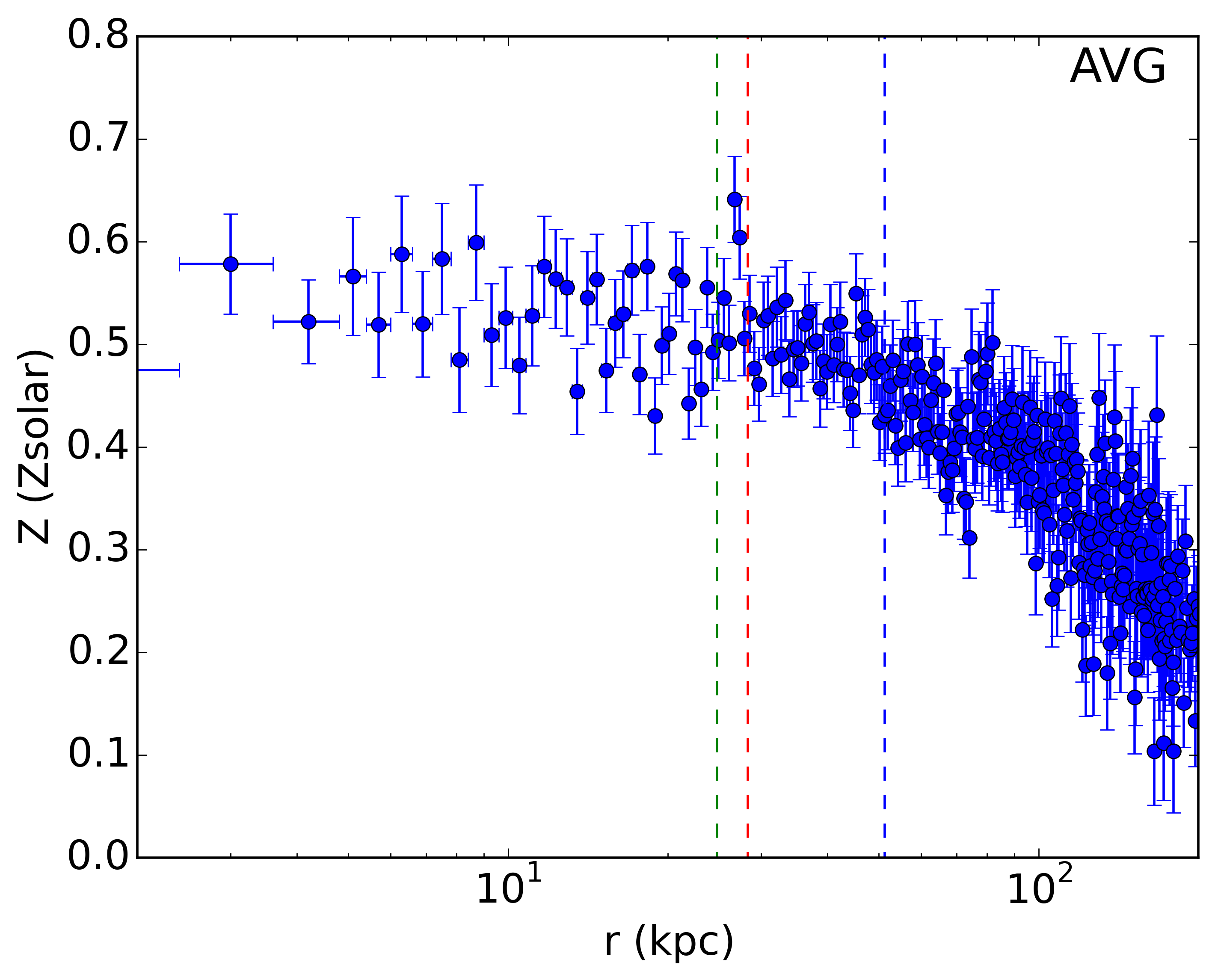} \\
\includegraphics[height=2.75in]{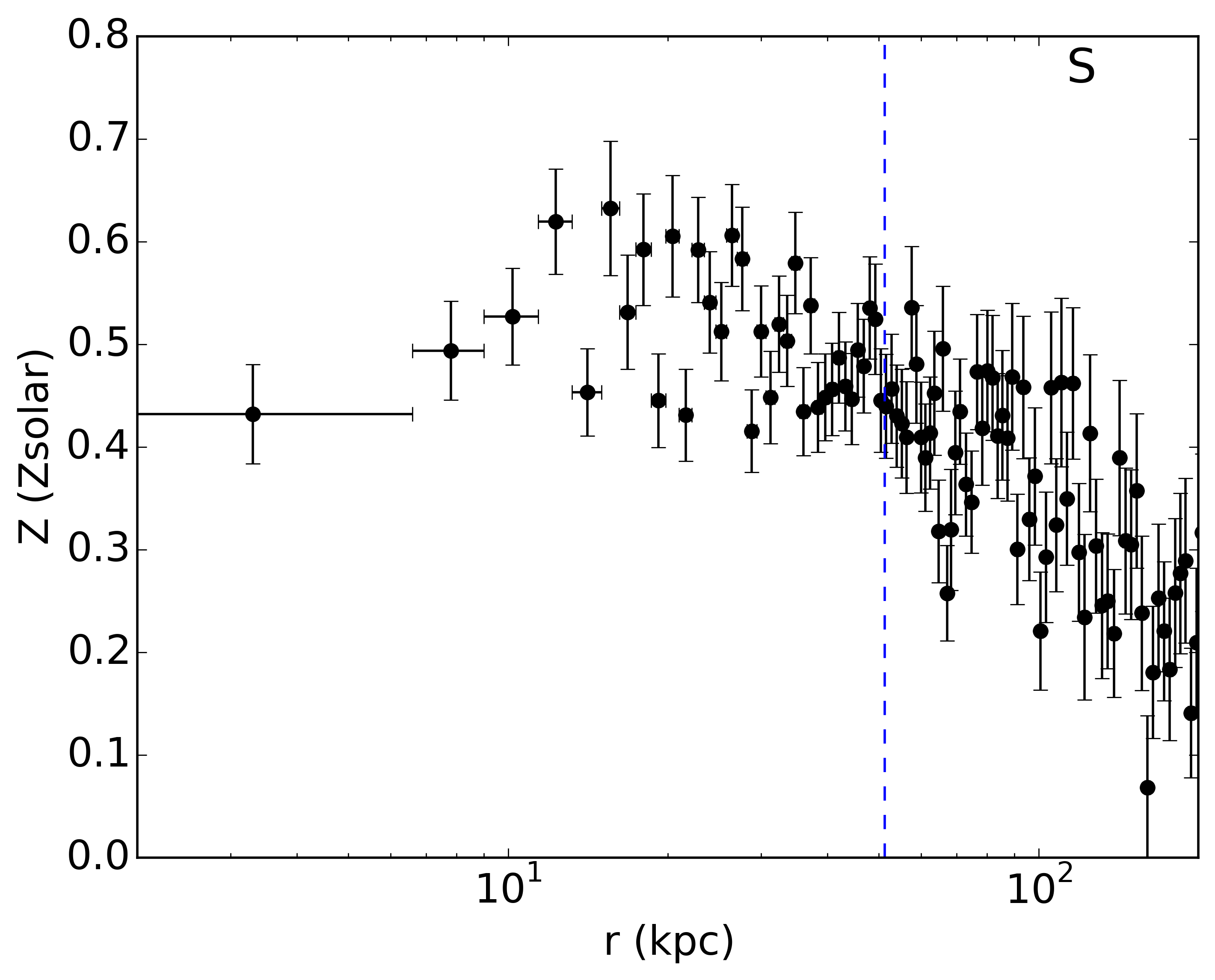} %
\includegraphics[height=2.75in]{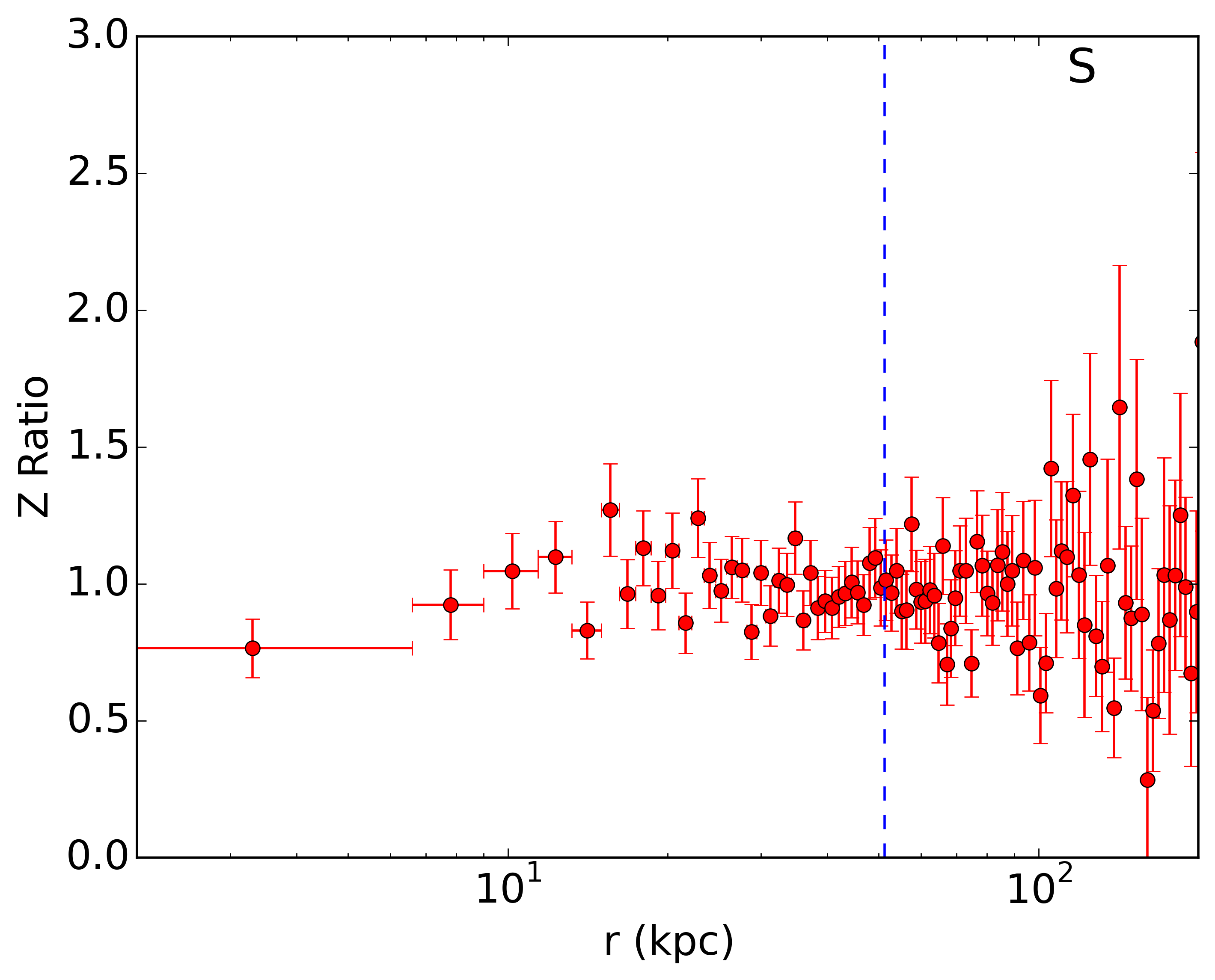} 
%% 1 column
% \includegraphics[height=2.0in]{profiles_x-ray/S_kT_kpc_big.png} 
% \includegraphics[height=2.0in]{profiles_x-ray/AVG_Z_kpc.png} \\
% \includegraphics[height=2.0in]{profiles_x-ray/S_Z_kpc.png} 
% \includegraphics[height=2.0in]{profiles_x-ray/S_Z_kpc_div_line.png} 
\end{center}
\caption{\small \textit{Top left:} Temperature radial profile along the S direction (black) and average temperature profile (blue).  
\textit{Top right:} Average metallicity profile.
\textit{Bottom left:} Metallicity profile along the S.  
\textit{Bottom right:} Ratio of the metallicity profile toward the S to the average metallicity profile.
The vertical blue line indicates the end of the tail as determined from the X-ray residual map (Fig. \ref{figP2:wedges}).
The red and green vertical lines in the average metallicity plot indicate the position of the NW and SW edges, respectively.
\label{figP2:radial_xray_s}}
\vspace{0.15in}
\end{figure*}

The southern tail feature has been studied extensively and a number of explanations have been proposed for its origin.
One possibility is that the tail resulted from an outflow of material triggered by the central SMBH.
An alternative interpretation states that the giant filament consists of cold gas from the central galaxy that has been stripped out by dynamical friction.
\cite{Fabian2001} suggest that the cold tail is due to a ``cooling wake'' produced as the central cluster galaxy oscillates through the core of the cluster.
In this model the X-ray bright ICM cooled around the motion of the BCG, which also gave rise to an H$\alpha$ filament \citep{Fabian2001, McDonald2009}.
\cite{Markevitch2001} argue that the BCG stays with the gravitational potential to within its observed small 150 km/s velocity relative to the cluster average \citep{OegerleHill1994}, while the relative motion between the BCG and the gas results from the sloshing of the gas.

The AGN origin has been seen as unlikely since no compelling signatures of AGN activity, such as X-ray cavities or radio emission, have been found to accompany the tail in its full extent.
While the stripping option requires a complicated scenario that most likely involves ram pressure stripping beyond the scales of the BCG \citep[see][]{Ehlert2015}, the cooling wake model is supported by the bulk of the observational evidence on the tail's dynamics. 
The gas motions in this filament have been found to be modest and not turbulent \citep{Crawford2005}.
Additionally, \cite{Hu1985} find that most of the optical filament that accompanies the X-ray tail shares the velocity of the cluster rather than that of the BCG.
This velocity configuration agrees well with the cooling scenario, and argues against the stripping model where a significant velocity gradient is expected.

The radio emission in our map at 235~MHz covers the inner parts of the tail up to $\sim$30~kpc from the core. 
It is evident from the radial profiles at 235~MHz (Fig. \ref{figP2:radio_profiles}) that the radio emission in the S is not as extended as in the NW and SW.
The lack of an extended, aged radio plasma structure clearly aligned with the main body of the X-ray tail argues against the possibility that the tail was created by the AGN activity. 
Similarly, our radio maps show that no low-frequency emission is associated with the hook feature and the small-scale X-ray surface brightness hole at the end of the tail discussed by \cite{Crawford2005} and \cite{Walker2014}. The lack of corresponding low-frequency radio emission for both the hook feature and the small depression further disfavors an AGN origin for the cold tail.

The temperature profile along the tail (Fig. \ref{figP2:radial_xray_s}) confirms that south of the core the overall temperature is significantly lower than the average, 
but does not demonstrate a significant gradient along the direction of the tail. 
A jump in temperature is only evident at the end of the tail at $\sim$45~kpc, after which the temperature increases rapidly. 
The lack of a significant temperature gradient along the tail suggests that there is little or no mixing between the gas in the tail and the surrounding ICM.
It also shows that there is no significant cooling by the gas in the tail over the time interval required for the tail to reach its current length.

Using the available temperature and pressure information, we have derived the cooling properties of the gas in the tail. 
We find no significant gradient in the cooling time along the tail and measure an average cooling time of $500 \pm 70$~Myr in a region encompassing the tail. 
This average is consistent with the age of $\sim$300 Myr we compute for the end of the tail using the velocity $+150$ km/s as measured by \cite{OegerleHill1994}. 
We note that obtaining the cooling time requires making assumptions about the geometry of the regions in order to derive the reprojected gas densities. 
These assumptions introduce a systematic uncertainty that is potentially larger than the quoted statistical uncertainty, but would not change the qualitative agreement between the two age estimates. 
Furthermore, we detect no strong metallicity gradient in the tail (Fig. \ref{figP2:radial_xray_s}), which might be expected if the tail consists of material that has been dragged from the central BCG. 
Thus, all the presented evidence argues in favor of the cooling wake hypothesis. 

\begin{figure}[t]
\begin{center}
\includegraphics[height=3.40in]{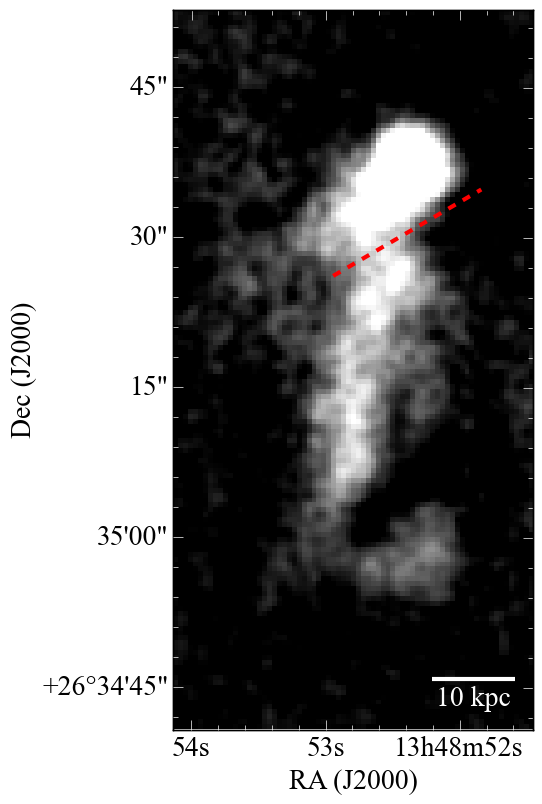}
\end{center}
\caption{\small Section of the X-ray residual map showing the cold tail. 
The dashed line shows the division between inner and outer tail as defined in the text.
\label{figP2:tail_res}}
\vspace{0.15in}
\end{figure}

Based on the X-ray residual and temperature maps, 
the tail can be divided into two main parts,  an inner round part and an outer straight part (Figs. \ref{figP2:temp_tail} and \ref{figP2:tail_res}).
The inner tail is separated from the outer tail by a bridge of warmer (3.5 keV) material situated at $\sim$6~kpc from the radio core. 
The inner tail is situated around the active core, it is not well confined, and is elongated toward the SE.
The outer part of the tail appears quite linear, but exhibits an inhomogeneous and knotty structure on smaller scales. 
The outer tail extends to the S out to a distance of 47 kpc from the center.

%\begin{figure}[!htbp]

Our interpretation of the different morphology of the inner and outer tail is that we are looking at a superposition of two major phenomena.
One is the dynamical movement of the central galaxy, which results in a cooling wake, and the other is related to the most recent episodes of AGN, which drives material out.
While the outer tail is governed by the movement of the BCG, the inner tail seems to be significantly affected by the AGN activity.
The inner tail follows the apparent NW--SE symmetry axis of enhanced pressure, presumably tracing the older AGN outflows.
We suggest that the AGN activity has likely disrupted the cooling wake, introducing the observed gap between the inner and outer tail, as well as the irregular shape of the inner tail and the kinks in the outer tail closer to the center.

%%%%%%%%%%%%%%%%%%%%%%%%%%%%%%%%%%%%%%%%%%%%%
%% DISCUSSION
%%%%%%%%%%%%%%%%%%%%%%%%%%%%%%%%%%%%%%%%%%%%%

\section{Discussion and conclusions}
\label{secP2:conclussions}

When revealed through shallow X-ray and high-frequency radio data, the morphology at the cores of relaxed cool-core clusters often looks simple and symmetric.
However, studying these objects with deeper and multiwavelength observations we find many complicated processes all acting in the same place.
In this work we presented extremely deep \textit{Chandra} X-ray data combined with GMRT radio observations at 235 MHz and 610 MHz.
We study A1795 as a dramatic example of a cluster core with rich X-ray morphology.
In this source the distinguished features are not all linked to a common origin.
In A1795 we are seeing motion of the core but also activity from the AGN.
These two phenomena act together to shape a very complicated picture, which makes it difficult to distinguish their individual effects.

Using a spatially resolved, X-ray spectral analysis, we identify the known X-ray depression in the NW direction and determine a distance of 44 kpc from the AGN. 
The inner rim of this depression is a bright edge of X-ray emission, situated 28 kpc from the core.
Due to its temperature structure, this elongated structure is consistent with a cold front.
Although no observed radio emission has previously been associated with the NW depression, our observation at 235 MHz shows that an extended filament of radio plasma connects the NW depression with the active core.
This configuration provides  strong circumstantial evidence that the surface brightness depression is a morphological signature of past AGN activity, and thus supports the classification of this depression as a cavity.
Based on the local sound speed we measure an age of 38 Myr for the cavity.
The power of the corresponding outburst is estimated to be $7\times10^{44}$erg/s.

The deep \textit{Chandra} data allowed us to identify two other cavities,  a SW cavity and an arc cavity.
Similarly to the case in the NW, in the SW direction we identified an edge (situated 25 kpc from the center) and an outer cavity fully covered by an extension of the radio plasma at 235 MHz.
For the SW cavity we computed an age of 30 Myr and a power of $1\times10^{44}$ erg/s.
Closer to the core we observed a very elongated arc-shaped cavity.
It is situated outside of the arc-like inner edge and is separated from the NW and the SW cavities by the NW and SW edges, respectively.
The arc cavity is located at a distance of 16 kpc NW from the center, which corresponds to an age of 14 Myr.
We compute a corresponding power of the AGN outburst of $9\times10^{43}$erg/s.
All three identified cavity structures are clearly distinguished as depressions in the radial surface brightness profiles and also align with peaks of hotter emission in the extracted temperature profiles.
While the cavities appear to be filled with hotter plasma, presumably injected by the AGN, they do not seem to have dredged up enhanced metallicity gas.

We interpret the complicated structure seen in the NW direction as the signature of three epochs of AGN activity. 
We see two edges: the NW edge and the arc edge, which is closer to the core.
Those two regions of compressed material serve as the boundaries between the three cavities corresponding to the three outbursts. 
The oldest is the NW cavity, then followed by the arc cavity. Inside the arc edge there are inner cavities, identified by \cite{Ehlert2015}, that are currently being inflated by the radio jets.
A similar picture is revealed in the SW; however, the signatures of AGN activity in this direction are less pronounced, most likely due to the effects of the moving core and the cooling wake.
Beyond the SW edge we observe the outer SW cavity, while closer to the core is the southern part of the arc cavity.
The elongated shape and the location of the arc cavity, shared between the NW and SW directions, may suggest that this irregular cavity is a result of the merger of two cavities produced by the same outburst. 

The observed radio behavior further supports the several-burst model.
The emission corresponding to the two radio extensions reaching beyond the edges appears very aged, exhibiting a spectral index between 235 and 610 MHz steeper than $-2.9$.
The spectral index profiles in radio support the notion that the extensions observed in the GMRT map at 235~MHz consist of aged plasma, presumably produced from an older outburst.
Moreover, the similar spectral properties of the two extensions suggest that they have similar ages and are thus likely produced by the same AGN outburst.

On the other hand, the spectral index of the radio emission remains relatively flat (flatter than $-1.0$) inside the two edges.
In this region, the new activity appears to be offsetting the ongoing spectral aging process.
The reacceleration process flattens the spectrum inside the edge, while the plasma outside of the edge is fully decoupled from the AGN and continues to age.

The new deep X-ray data allows us to separate the $\sim$47 kpc  tail into two main parts with very different structures.
While the inner part is round, the outer tail appears straight and very confined.
Based on the different morphology at the two parts of the tail, we suggest that the outer tail is governed by the movement of the BCG, while the inner tail is severely disturbed by the AGN activity.
In this scenario, the observed gap between the inner and outer tail, the irregular shape of the inner tail, and the kinks observed in the outer tail are all attributed to the activity of the AGN.

Even at 235 MHz we do not observe radio emission corresponding to the extended cold tail.
The lack of this emission disfavors an AGN origin of the tail.
Furthermore, despite the depth of the X-ray data, we detect no significant temperature, cooling time, or metallicity gradient along the tail.
We measure an average cooling time of $500 \pm 70$ ~Myr, which is consistent with the age of $\sim$300 Myr derived from the velocity of the core movement \citep{OegerleHill1994}. 
All of the present evidence supports the cooling wake hypothesis for the origin of the tail.

\cite{Giacintucci2014} have identified a mini-halo in the system and determine its size and power based on a VLA observation taken with the C-array configuration. 
They estimate a size for the mini-halo of $\sim$100 kpc. 
Although the resolution of the GMRT images presented here is higher than the images of \cite{Giacintucci2014}, we do not resolve the region of the mini-halo inside a radius of $\sim$20~kpc from the radio core. 
However, beyond $\sim$30~kpc, we observe signatures of AGN activity which are consistent with the X-ray data and are supported by the spectral index derived between 235 and 610 MHz. 
Since our analysis interprets some of the extended radio emission to be a result of AGN activity, the implication is that the mini-halo is smaller in size and power than originally inferred by \cite{Giacintucci2014}.

In this work, we have presented evidence for the signatures of multiple episodes of AGN activity in the core of the A1795 cluster. 
Despite the use of very deep X-ray and broadband radio data, it is still difficult to disentangle the observed morphology of the core into a simple feedback-driven model. 
The ongoing AGN activity is clearly superimposed on core structures related to the motion of the central BCG, the apparent sloshing of the ICM gas in the central potential, and the formation of the cooling wake these motions have induced. 
In the radio, the observed morphology is similarly complicated, due to the presence of what appears to be aged relic emission from past AGN activity mingled with an apparent radio mini-halo in the core. 
It is possible that we are looking at a mini-halo being formed by multiple episodes of activity and the complicated picture revealed in the core of A1795 may ultimately provide evidence of the mechanism responsible for building mini-halos. 
Future higher resolution, low-frequency radio data from LOFAR and gas velocity information provided by future missions such as Athena may allow us to separate these overlapping physical effects and better constrain the evolution of the observed radio and X-ray structures in the core of A1795 over time.

%%%%%%%%%%%%%%%%%%%%%%%%%%%%%%%%%%%%%%%%%%%%%
%% ================================================================
%%%%%%%%%%%%%%%%%%%%%%%%%%%%%%%%%%%%%%%%%%%%%

\begin{acknowledgements}
    
We thank Laura B\^irzan and Simona Giacintucci for kindly providing their VLA images at 1.4 GHz. 
GDK acknowledges support from NOVA (Nederlandse Onderzoekschool voor Astronomie).
GMRT is run by the National Centre for Radio Astrophysics of the Tata Institute of Fundamental Research.
This research has made use of the NASA/IPAC Extragalactic Database (NED) which is operated by the Jet Propulsion Laboratory, California Institute of Technology, under contract with the National Aeronautics and Space Administration. 
We have also used SAOImage DS9, developed by Smithsonian Astrophysical Observatory.       
      
\end{acknowledgements}

\bibliographystyle{aa} 
\bibliography{bibtex/refs_GDK_PhD_02}

\end{document}